\documentclass[aps,prb,twocolumn,superscriptaddress,showpacs]{revtex4-1}
\usepackage{amsmath}
\usepackage{color}
\usepackage{amssymb}
\usepackage{epsfig}
\usepackage{multirow}
\usepackage{amsbsy}
\usepackage{array}
\usepackage{diagbox}
\usepackage{bm}
\usepackage{extarrows}
\usepackage{graphicx}
\usepackage{appendix}

\usepackage[colorlinks=true,linkcolor=blue,citecolor=blue, urlcolor=blue,bookmarks=false]{hyperref}
\begin{document}

\title{Classical and quantum order in hyperkagome antiferromagnets}

\author{Hui-Ke Jin}
\affiliation{Department of Physics, Zhejiang University, Hangzhou 310027, China}

\author{Yi Zhou}
\email{yizhou@iphy.ac.cn}
\affiliation {Beijing National Laboratory for Condensed Matter Physics and Institute of Physics, Chinese Academy of Sciences, Beijing 100190, China}
\affiliation{Department of Physics, Zhejiang University, Hangzhou 310027, China}
\affiliation{Kavli Institute for Theoretical Sciences and CAS Center for Excellence in Topological Quantum Computation, University of Chinese Academy of Sciences, Beijing 100190, China}
\affiliation{Songshan Lake Materials Laboratory, Dongguan, Guangdong 523808, China}

\date{\today}

\begin{abstract}
	Motivated by recent experiments and density functional theory calculations on choloalite PbCuTe$_2$O$_6$, which possesses a Cu-based three-dimensional hyperkagome lattice, we propose and study a $J_1$-$J_2$-$J_3$ antiferromagnetic Heisenberg model on a hyperkagome lattice. In the classical limit, possible ground states are analyzed by two triangle rules, i.e., the ``hyperkagome triangle rule" and the ``isolated triangle rule," and classical Monte Carlo simulations are exploited to identify possible classical magnetic ordering and explore the phase diagram. In the quantum regime, Schwinger boson theory is applied to study possible quantum spin liquid states and long-range magnetically ordered states on an equal footing. These quantum states with bosonic partons are classified and analyzed by using projective symmetry groups (PSGs). It is found that there are only four types of algebraic PSGs allowed by the space group $P4_{1}32$ on a hyperkagome lattice. Moreover, there are only two types of PSGs that are compatible with the $J_1$-$J_2$-$J_3$ Heisenberg model. These two types of $Z_2$ bosonic states are distinguished by the gauge-invariant flux on the elementary ten-site loops on the hyperkagome network, called zero-flux state and $\pi$-flux state respectively. Both the zero-flux state and the $\pi$-flux state are able to give rise to quantum spin liquid states as well as magnetically ordered states, and the zero-flux states and the $\pi$-flux states can be distinguished by the lower and upper edges of the spectral function $S(\bm{q},\omega)$, which can be measured by inelastic neutron scattering experiments.
	
\end{abstract}

\maketitle


\section{Introduction}

{\em Magnetism in frustrated lattices.---}
Before the phrase geometric frustration (or frustration for short) was first introduced by Toulouse in the context of spin glass in 1970s~\cite{Toulouse77,Toulouse77a}, frustrated magnetic systems had been long studied indeed. Recently, it has been becoming an important feature in magnetism~\cite{BookDiep,BookLacroix}. In an early paper, Wannier~\cite{wannier} studied the classical Ising model on a triangular lattice with the antiferromagnetically coupled nearest neighbor (NN) spins, which is known as the simplest example of geometric frustration now~\cite{BookDiep}. The frustration on a triangular lattice gives rise to numerous ground-state degeneracies in this model. Starting with one minimum-energy spin configuration, one is able to obtain another minimum-energy configuration by a ``local" distortion. As a result, it does not order magnetically even at zero temperature due to huge residual entropy~\cite{wannier}, and the spin-spin correlation has been found to decay following a power law at zero temperature in accordance with the exact solution~\cite{Stephenson70}. Such a disordered state is also called a classical spin liquid state. However, the minimum-energy spin configurations for the classical antiferromagnetic (AFM) Heisenberg model on a triangular lattice are not locally degenerate. They are distinguished from each other only by a ``global" spin $SO(3)$ rotation. Consequently, the spins form a 120$^{\circ}$ long-range magnetic order at zero temperature.

Indeed, a lattice can be ``more frustrating" than a triangular lattice, such that the classical NN AFM Heisenberg model also has an infinite number of degenerate ground states that are connected to one another by continuous ``local" distortions of the spin configuration~\cite{orderbyd1}. This property holds on any lattice with ``corner-sharing" units, such as checkerboard, kagome, and pyrochlore lattices~\cite{Moessner98}. As an example, we shall illustrate this property on a three-dimensional (3D) hyperkagome lattice in Sec.~\ref{sec:hyperklimit} in this paper. 

In addition to mentioned magnetically ordered or disordered ground states and possible spin glass state~\cite{Toulouse77,Toulouse77a}, many other emergent phenomena have been widely studied in classical spin systems on frustrated lattices as well, such as spin ice~\cite{spinice} and the effect of order by disorder~\cite{orderbyd1,orderbyd2}.

{\em Effects of quantum fluctuations.---} Quantum mechanics brings new features to frustrated spin systems. (i) On one hand, quantum fluctuations will lift the local ground state-degeneracy in the classical model and lead to a ``classical" ordered state, e.g., the 120$^{\circ}$ magnetic ordering ground state in the NN AFM Heisenberg model on a triangular lattice. This classical order is associated with spontaneous symmetry breaking and is characterized by the long-range correlation of ``local" operators, say, the local spin operators in this situation. (ii) On the other hand, strong quantum fluctuations that are enhanced in frustrated lattices may destroy the long-range magnetic ordering and give rise to a quantum spin liquid ground state~\cite{QSLs1,QSLs2}. It is different from a classical spin liquid state due to the residual entropy that the entropy density in a quantum spin liquid state is zero. This difference serves as one of the criteria in the experimental identification of quantum spin liquids and can be examined by specific heat measurements.

As a combination of the two effects, say, the lift of the local degeneracy and the ruin of classical magnetic order, quantum mechanics may lead to exotic ground states and low-energy behaviors on frustrated lattices that cannot be captured by traditional semiclassical approaches. This inspired people to study novel quantum states in frustrated lattices, especially quantum spin liquid states~\cite{QSLs1,QSLs2}. These quantum states are not associated with the classical order described by Landau's symmetry-breaking paradigm. Instead, a concept of ``quantum order" was proposed to describe the non-symmetry-breaking orders that appear in such quantum states generally~\cite{PSG1,PSG2}.

{\em Quantum order and projective symmetry group.---}
The quantum order generalizes the topological order of gapful states (with long-range entanglement)~\cite{WenRMP17} to gapless states. To describe the quantum order in quantum spin liquid states, a new mathematical object---projective symmetry group (PSG)---was introduced~\cite{PSG1}. The concept of quantum order and its PSG characterization allow us to classify symmetric quantum spin liquid states and understand the quantum phase transitions between them in a systematic way. A quantum spin liquid state is characterized by fractionalized spinons and accompanied gauge fields. The spinons can be either fermionic as Abrikosov fermions~\cite{Abrikosov65,BZA,Baskaran88} or bosonic as Schwinger bosons~\cite{largeN1}. Mathematically, it is possible to rewrite a quantum spin operator in terms of parton (fermionic or bosonic spinon) operators. Thus, the Hilbert space is enlarged and an additional gauge structure is generated. To obtain physical spin states, the Gutzwiller projection is employed to remove the gauge redundancy and restore the physical Hilbert space. Note that the gauge structure of the low-energy effective theory of partons could be different from the largest gauge redundancy subject to the Gutzwiller projection. The PSG is to characterize fractionalized parton states with additional gauge structures, which is valid for both fermionic and bosonic spinons~\cite{QSLs1}. In particular, the Schwinger boson approach to quantum spins, which we will utilize in this paper, is able to describe either a quantum spin liquid with a finite spinon excitation gap or a magnetically ordered state when the spinon gap closes and Bose-Einstein condensation of spinons occurs~\cite{MichaelMa89}. It means that that the Schwinger boson approach allows us to treat gapped spin liquid states and magnetically ordered states on an equal footing.

{\em Realistic materials.---} In the past two decades, a host of AFM insulators have been discovered in various frustrated lattices, and some of them suggest quantum spin liquid ground states. The candidate materials include  organic salts $\kappa$-(ET)$_{2}$Cu$_{2}$(CN)$_{3}$~\cite{organic1} and EtMe$_{3}$Sb[(Pd(dmit)$_{2}$]$_{2}$~\cite{organicdmit} on anisotropic triangular lattice, rare-earth compound YbMgGaO$_4$~\cite{YbMg1,YbMg2,YbMg3} on isotropic triangular lattice, Herbertsmithite ZnCu$_{3}$(OH)$_{6}$Cl$_{2}$~\cite{Kagome07} and Zn-doped Barlowite Cu$_3$Zn(OH)$_6$FBr~\cite{barlowite15,barlowite17} on kagome lattice, and spinel oxide Na$_{4}$Ir$_3$O$_8$~\cite{NaIrO1,NaIrO1.2,NaIrO1.3} on three-dimensional (3D) hyperkagome lattice. None of these materials exhibit classical magnetic order down to the lowest low temperature. 

{\em Hyperkagome Na$_4$Ir$_3$O$_8$.--- }
Among these candidate materials, we are particularly interested in the spinel compound Na$_{4}$Ir$_3$O$_8$, in which the $S=1/2$ spins come from the $5d^5$ Ir$^{4+}$ ions and form a solid network of corner-sharing triangles, i.e., a 3D hyperkagome lattice. The isotropic NN AFM Heisenberg model was proposed as a microscopic spin model for this compound in several works~\cite{NaIrO2,NaIrO3,NaIrO4,NaIrO6}. Highly degenerate classical ground states were numerically found in Ref.~[\onlinecite{NaIrO3}], in which the spins are treated as classical $N$-component vectors and the model is of $O(N)$ rotational symmetry. With the help of Schwinger bosons, a coplanar magnetically ordered ground state was predicted in the semiclassical limit, while a gapped topological $Z_2$ spin liquid was proposed in the quantum limit~\cite{NaIrO6}. On the other hand, gapless spin liquid states with fermionic spinons were suggested~\cite{NaIrO2, NaIrO4}, which are in good agreement with magnetic susceptibility and thermodynamic measurements~\cite{NaIrO1}. The effect of strong spin-orbit coupling due to the large atomic number of Ir was also addressed~\cite{NaIrO5} and was used to explain the anomalously large Wilson ratio observed in experiments~\cite{NaIrO4,ChenKim13}.

{\em Hyperkagome PbCuTe$_2$O$_6$.--- }
More recently, another hyperkagome lattice compound PbCuTe$_2$O$_6$ was synthesized~\cite{PbCuTe1} and suggested as a candidate for quantum spin liquid~\cite{PbCuTe2}, where spins come from Cu$^{2+}$ ($3d^{9}$) ions and crystallize a Cu-based hyperkagome network. It is different from Na$_4$Ir$_3$O$_8$, where Ir atoms are much larger than sodium and oxygen, that the $5p$ orbitals in Te atoms are extensive in the choloalite compound PbCuTe$_2$O$_6$.  So that the isolated triangles are geometrically smaller than the hyperkagome triangles. We define the ordering of neighbors in accordance with their bond lengths rather than the strengths.
Consequently, the first NN ($J_1=J_{tri}$) bonds form isolated triangles, the second NN ($J_2=J_{hyper}$) bonds form a hyperkagome lattices, while the third NN ($J_3=J_{chain}$) bonds form uniform chains passing along the three crystallographic directions. All these three types of bonds are AFM and considerable, although the AFM coupling on the hyperkagome lattice is the largest on, i.e., $J_{tri}/J_{hyper}\approx 0.54$ and $J_{chain}/J_{hyper}\approx 0.77$, in accordance with density functional theory (DFT) calculation on hopping integrals~\cite{PbCuTe1}.  
Both thermodynamic measurements~\cite{PbCuTe1} and NMR and muon spin relaxation studies~\cite{PbCuTe2} suggest the absence of magnetic order in polycrystalline PbCuTe$_2$O$_6$ sample and a gapless spin liquid state with fermionic spinons.

Motivated by experimental observations and DFT calculations on PbCuTe$_2$O$_6$, we study a $J_1$-$J_2$-$J_3$ Heisenberg model and possible classical and quantum orders in hyperkagome antiferromagnets in this paper.

The rest of this paper is organized as follows. In Sec.~\ref{sec:symmetry}, we introduce the space group and lattice structure, and present a $J_{1}$-$J_{2}$-$J_{3}$ Heisenberg model for PbCuTe$_2$O$_6$. 
In Sec.~\ref{sec:corder}, we analyze possible classical spin orders for the proposed $J_{1}$-$J_{2}$-$J_{3}$ Heisenberg model, and explore the phase diagram by Monte Carlo simulation.
In Sec.~\ref{sec:sboson}, we formulate Schwinger boson theory and classify bosonic states with the help of PSGs.
In Sec.~\ref{sec:results}, possible long-range magnetically ordered states and quantum spin liquid states are analyzed in accordance with PSGs and Schwinger boson mean-field theory.
Sec.~\ref{sec:summary} is devoted to a summary.

\section{Symmetry, lattice and model Hamiltonian}\label{sec:symmetry}

We use the material PbCuTe$_2$O$_6$ as an example to illustrate space group $P4_132$ (no.213) and corresponding lattice structure.

{\em Space group $P4_{1}32$.--- } There are two types of hyperkagome lattices corresponding to non-symmorphic space groups $P4_132$ and $P4_332$, which differ in chirality only~\cite{grouptable}. For simplicity, we will focus on $P4_132$ in this paper. The hyperkagome lattice is a cubic lattice generated by lattice translations $T_{1,2,3}$ for cubic unit cell along three directions,
\begin{equation}\label{eq:T123}
\begin{array}{l}
(x, y, z) \xrightarrow{T_{1}} (x+1, y, z), \\
(x, y, z) \xrightarrow{T_{2}} (x, y+1, z), \\
(x, y, z) \xrightarrow{T_{3}} (x, y, z+1), \\
\end{array}
\end{equation}
where we set lattice constant as unit for simplicity. The corresponding point group is the octahedral group $O$ consisting of 24 symmetry operations~\cite{grouptable}. In such a nonsymmorphic space group, the fourfold rotation in the octahedral group $O$ is replaced by the fourfold screw operation. One of the non-symmorphic screw rotations, $S_4$, is given by a 90$^{\circ}$ rotation along the axis $(x, -1/4, 1/2)$ followed by a fractional translation of $(1/4, 0, 0)$,
\begin{equation}
(x, y, z)  \xrightarrow{S_{4}} (\frac{1}{4}+x, \frac{1}{4}-z, \frac{3}{4}+y). \label{eq:S4}
\end{equation}
The four times of the fourfold screw operations result in a displacement by a lattice constant, say, $(S_{4})^{4}=T_{1}$.

In addition to these two symmetry operations, there are two other symmetry operations which generate the whole space group $P4_{1}32$ together with $T_{1}$ and $S_4$.
One is the twofold rotation, $C_2$, along $(3/8, 3/4-y, y)$ axis,
\begin{equation}
(x, y, z)  \xrightarrow{C_{2}} (\frac{3}{4}-x, \frac{3}{4}-z, \frac{3}{4}-y). \label{eq:C2}
\end{equation}
The other is the threefold rotation along $(1,1,1)$ direction,
\begin{equation}
(x, y, z)  \xrightarrow{C_{3}} (z, x, y). \label{eq:C3}
\end{equation}
The commutation relations of these symmetry operations are given in Appendix~\ref{ap:PSG}.

\begin{figure}[tbp]
	\centering
	\includegraphics[width=8.4cm]{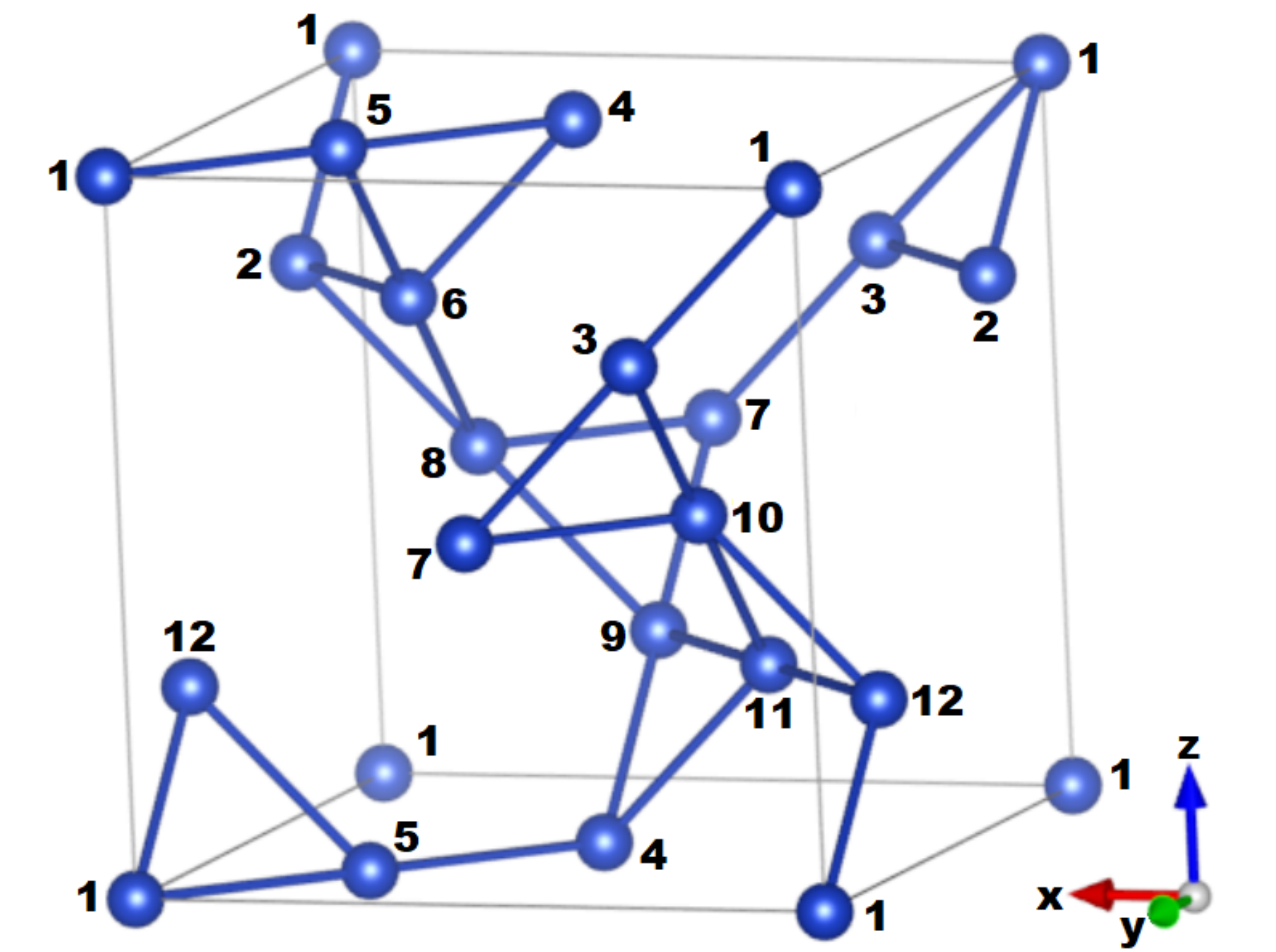}
	\caption{A unit cell in a hyperkagome lattice, which is formed by a 3D network of corner-sharing triangles. The edges of these triangles are called ``hyperkagome bonds," which are also defined as second nearest neighboring (NN) bonds in the main text. The 12 sublattices are labeled by $\mu=1,2,\cdots,12$.}
	\label{fig:hpkgmlatt}
\end{figure}

{\em Lattice structure.---} The 24 symmetry operations generate 24$e$ general positions, among which there are 12$d$ special positions with higher symmetry. A hyperkagome lattice is made of these 12$d$ special positions, which is occupied by Ir atoms in Na$_4$Ir$_3$O$_8$ and Cu atoms in PbCuTe$_2$O$_6$, such that each unit cell consists of 12 lattice sites. These lattice sites can be labeled by a unit cell index $i$ and a sublattice index $\mu=1,2,\cdots,12$ (see Fig.~\ref{fig:hpkgmlatt}). 
All the sublattice coordinates can be found in Table~\ref{tab:12d}.
\begin{table}[tbp]
\caption{12$d$ special positions of space group $P4_{1}32$}\label{tab:12d}
\renewcommand\arraystretch{1.5}
\begin{tabular}{|l|l|l|}
	\hline
	1 (${}{\frac{3}{8}, \bar{y}, {y}+\frac{3}{4}}$)  & 2 (${}{\bar{y}+\frac{1}{2}, \bar{y}+\frac{3}{4}, \frac{5}{8}}$)  & 3 (${}{\bar{y}+\frac{1}{4}, \frac{7}{8}, {y}+\frac{1}{2}}$) \\ \hline
	4 (${}{{y}+\frac{3}{4}, \frac{3}{8}, \bar{y}}$) &
	5(${}{{y}+\frac{1}{2}, \bar{y}+\frac{1}{4}, \frac{7}{8}}$ ) & 6(${}{\frac{5}{8}, \bar{y}+\frac{1}{2}, \bar{y}+\frac{3}{4}}$) \\ \hline
	7 ( ${}{\bar{y}, {y}+\frac{3}{4}, \frac{3}{8}}$) & 8 (${}{\bar{y}+\frac{3}{4}, \frac{5}{8}, \bar{y}+\frac{1}{2}}$ ) & 9 (${}{\frac{7}{8}, y+\frac{1}{2}, \bar{y}+\frac{1}{4}}$ )  \\ \hline
	10 (${}{\frac{1}{8}, {y}, {y}+\frac{1}{4}}$)     & 11 (${}{{y}, {y}+\frac{1}{4}, \frac{1}{8}}$ )                    & 12 (${}{{y}+\frac{1}{4}, \frac{1}{8}, {y}}$)                \\
	\hline
\end{tabular}
\end{table}
Note that the 12$d$ positions (sublattices) form a closed set  $S_4$ screw and $C_2$ and $C_3$ rotations, which are summarized in Table~\ref{tab:aor}.
\begin{table}[tbp]
	\caption{Actions of screw and rotations. A sublattice $(x,y,z;\mu)$ is transferred to $(x^{\prime},y^{\prime},z^{\prime};\mu^{\prime})$ under screw $S_4$ and rotations $C_2$ and $C_3$. Here $(x,y,z)$ are the coordinates of a sublattice $\mu$.}\label{tab:aor}
	\renewcommand\arraystretch{1.5}
	\begin{tabular}{|l|l|l|l|}
		\hline
		$\mu$ & $C_{2}$                            & $C_{3}$         & $S_{4}$                    \\
		\hline
		1     & $(\bar{x}, \bar{z}, \bar{y}; 1)$   & $(z, x, y; 4)$  & $(x, \bar{z}-1, y; 6)$     \\
		\hline
		2     & $(\bar{x}, \bar{z}, \bar{y}; 12)$  & $(z, x, y; 6)$  & $(x, \bar{z}-1, y+1; 8)$   \\
		\hline
		3     & $(\bar{x}, \bar{z}, \bar{y}-1; 5)$ & $(z, x, y; 5)$  & $(x, \bar{z}-1, y+1; 2)$   \\
		\hline
		4     & $(\bar{x}, \bar{z}, \bar{y}; 7)$   & $(z, x, y; 7)$  & $(x+1, \bar{z}, y+1; 11)$  \\
		\hline
		5     & $(\bar{x}, \bar{z}-1, \bar{y}; 3)$ & $(z, x, y; 9)$  & $(x, \bar{z}-1, y+1; 4)$   \\
		\hline
		6     & $(\bar{x}, \bar{z}, \bar{y}; 10)$  & $(z, x, y; 8)$  & $(x, \bar{z}-1, y+1; 9)$   \\
		\hline
		7     & $(\bar{x}, \bar{z}, \bar{y}; 4)$   & $(z, x, y; 1)$  & $(x, \bar{z}-1, y+1; 3)$   \\
		\hline
		8     & $(\bar{x}, \bar{z}; \bar{y}, 11)$  & $(z, x, y; 2)$  & $(x+1, \bar{z}-1, y+1; 7)$ \\
		\hline
		9     & $(\bar{x}-1, \bar{z}, \bar{y}; 9)$ & $(z, x, y; 3)$  & $(x+1, \bar{z}, y+1; 10)$  \\
		\hline
		10    & $(\bar{x}, \bar{z}, \bar{y}; 6)$   & $(z, x, y; 12)$ & $(x, \bar{z}, y; 1)$       \\
		\hline
		11    & $(\bar{x}, \bar{z}, \bar{y}; 8)$   & $(z, x, y; 10)$ & $(x, \bar{z}, y+1; 12)$    \\
		\hline
		12    & $(\bar{x}, \bar{z}, \bar{y}; 2)$   & $(z, x, y; 11)$ & $(x, \bar{z}, y; 5)$       \\
		\hline
	\end{tabular}
\end{table}

\begin{figure}[tbp]
	\centering
	\includegraphics[width=8.4cm]{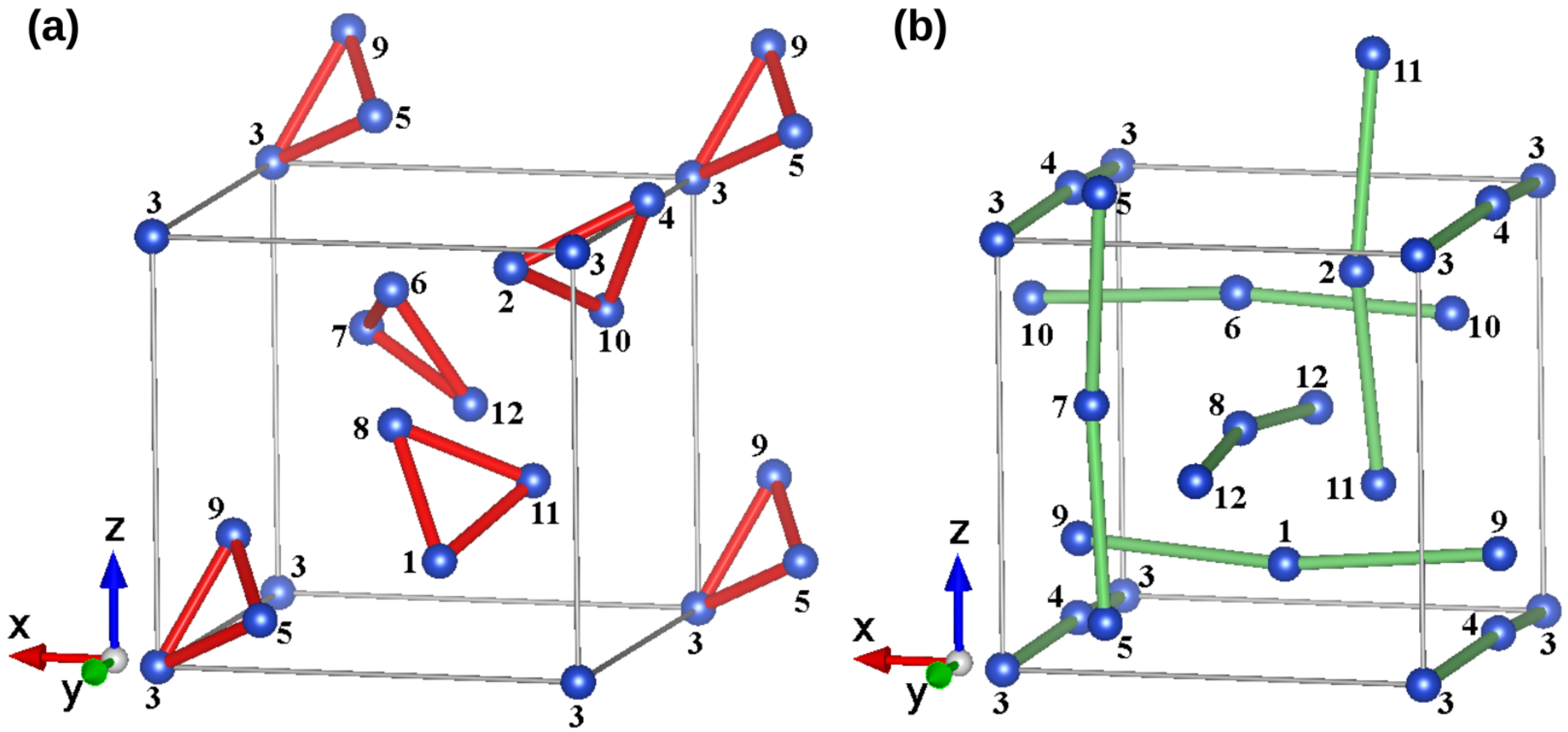}
	\caption{(a) Four types of isolated triangles, whose edges are called ``for isolated-triangle bonds" or first NN bonds. (b) Six types of uniform chains, which pass along all the three  crystallographic directions. The links on these uniform chains are called ``chain bonds" or third NN bonds. }
	\label{fig:triangle_chain}
\end{figure}

{\em Triangles and chains.---} As illustrated in Fig.~\ref{fig:hpkgmlatt}, there are eight types of corner-sharing triangles on a hyperkagome lattice, whose edges give rise to a connect 3D network and are called ``hyperkagome bonds." We also call these corner-sharing triangles as ``hyperkagome triangles." These hyperkagome triangles can be characterized by a triad of sites $(\mu=1,2,\cdots,12)$ as follows:
\begin{subequations}\label{eq:tri.chain}
\begin{equation}
\begin{array}{llll}
\lbrack 1,2,3],  & [4,6,5], & [7,8,9],  & [10,11,12], \\
\lbrack 1,5,12], & [2,6,8], & [3,7,10], & [4,11,9].
\end{array}
\end{equation}
Here we neglect unit cell index $i$ for simplicity.

To model magnetic interactions for PbCuTe$_2$O$_6$, we also need to define four types of isolated triangles [see Fig.~\ref{fig:triangle_chain}(a)] given by the following triads of sites,
\begin{equation}
[1, 8, 11],\,\, [2, 4, 10],\,\, [3, 5, 9],\,\, [6, 7, 12],
\end{equation}
and six types of uniform chains [see Fig.~\ref{fig:triangle_chain}(b)] given by the following linkings:
\begin{equation}\label{eq:chain}
\begin{array}{lll}
\lbrack 1, 9, 1, 9], & [2, 11, 2, 11], & [3, 4, 3, 4],   \\
\lbrack 5, 7, 5, 7], & [6, 10, 6, 10], & [8, 12, 8, 12],
\end{array}
\end{equation}
\end{subequations}
which pass along all the three crystallographic directions. These isolated triangles and uniform chains are illustrated in Fig.~\ref{fig:triangle_chain}.

{\em Model Hamiltonian.---} PbCuTe$_2$O$_6$ compounds are magnetic Mott insulators, where spins $S=1/2$ come from Cu$^{2+}$~($3d^9$) cations. The dominant magnetic exchange interaction between two Cu atoms is via Cu-O-Te-O-Cu paths, namely, hyperkagome bonds. There exist a strong hybridization between copper $d_{x^2-y^2}$ orbitals with oxygen $p_x$ and $p_y$ orbitals via $\sigma$ bonding and also with Te$^{4+}$~($5s^2$) cations, resulting in the dominant AFM exchange on hyperkagome bonds, $J_{hyper}>0$. 
Additionally, the AFM exchange on isolated triangle bonds, $J_{tri}>0$, and the AFM exchange along uniform chains, $J_{chain}>0$, are considerable too. On the other hand, in accordance with the distances between two Cu atoms, first nearest neighboring (NN) bonds are isolated triangle bonds, second NN bonds are hyperkagome bonds, and third NN bonds are along the uniform chains.

Taking into account all the magnetic interactions as discussed in the above, we model PbCuTe$_2$O$_6$ by the following $J_{1}$-$J_{2}$-$J_{3}$ Heisenberg model on a hyperkagome lattice,
\begin{equation}\label{eq:HJ123}
\mathcal{H}=\sum_{a=1}^{3}\sum_{\langle{}i\mu,j\nu\rangle_{a}}J_{a}\bm{S}_{i\mu}\cdot\bm{S}_{j\nu},
\end{equation}
where $\langle{}i\mu,j\nu\rangle_{a}$ denotes $a$th NN bonds and $\bm{S}_{i\mu}$ is the vector of spin operators at site $i\mu$.
All the couplings are AFM and are given by $J_1=J_{tri}$, $J_2=J_{hyper}$, and $J_3=J_{chain}$. In this paper, we shall study the $J_{1}$-$J_{2}$-$J_{3}$ Heisenberg model by using classical Monte Carlo method as well as Schwinger boson mean-field theory.

\section{Classical Magnetic Orders}\label{sec:corder}

In this section, we study possible ground states for the $J_{1}$-$J_{2}$-$J_{3}$ AFM Heisenberg model on a hyperkagome lattice defined in Eq.~\eqref{eq:HJ123} in the classical limit. We begin with analyzing degenerate ground states in the hyperkagome limit, say, $J_{tri}=J_{chain}=0$ and $J_{hyper}>0$, and then consider the couplings on isolated triangles $J_{tri}>0$ and the couplings on the uniform chains $J_{chain}>0$. Classical Monte Carlo simulations are performed to explore the phase diagram.

\subsection{Hyperkagome limit $J_{tri}=J_{chain}=0$: Local degeneracy of ground states}\label{sec:hyperklimit}

\begin{figure}[tbp]
	\centering
	\includegraphics[width=3.6cm]{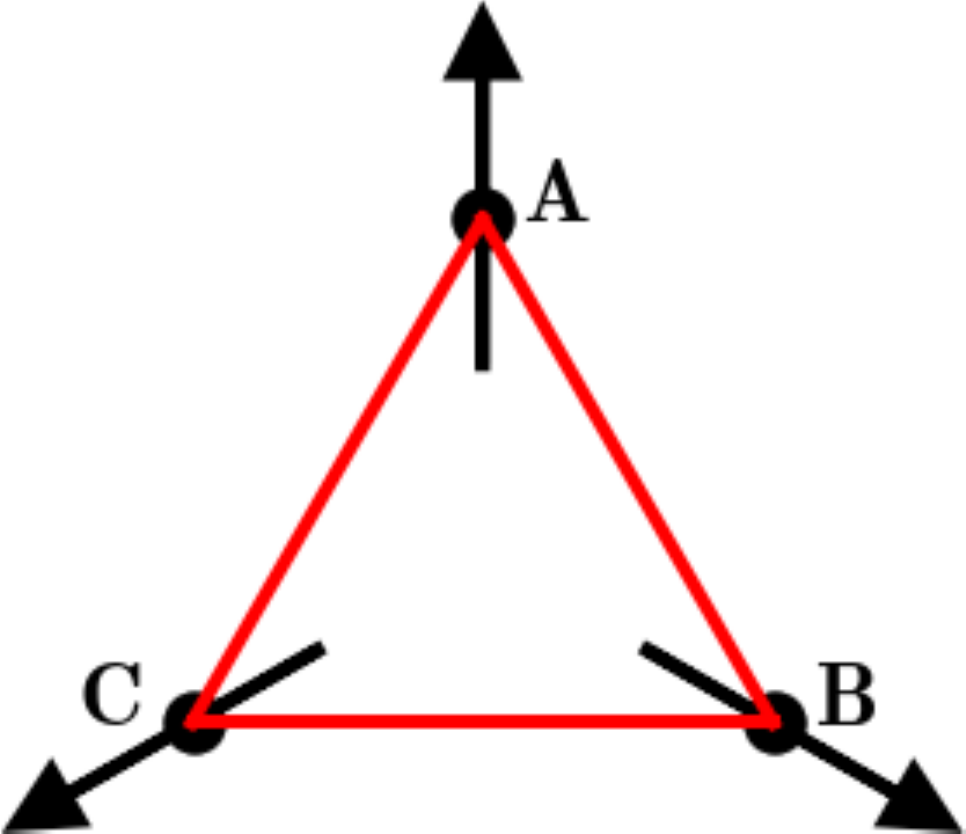}
	\caption{Three spin vectors $\bf{A}$, $\bf{B}$, and $\bf{C}$ pointing toward the vertices of an equilateral triangle. }
	\label{fig:120neel}
\end{figure}

In the hyperkagome limit $J_{tri}=J_{chain}=0$, the only active magnetic coupling is $J_{hyper}>0$ on hyperkagome bonds. Similar to the 2D kagome lattice, the hyperkagome lattice consists of corner-shared triangles which form a 3D network. Then the Hamiltonian in Eq.~\eqref{eq:HJ123} can be simplified as the sum of the squares of the total spins $\vec{S }_{\bigtriangleup }=\vec{S}_{1}+\vec{S}_{2}+\vec{S}_{3}$
on individual hyperkagome triangles,
\begin{equation}\label{eq:HJ2}
\mathcal{H}=J_{hyper}\sum_{\bigtriangleup }(\vec{S}_{\bigtriangleup })^{2}.
\end{equation}
Classical ground states are obtained whenever
\begin{equation}\label{eq:hyper_rule}
\vec{S}_{\bigtriangleup }=0.
\end{equation}
The {\it hyperkagome triangle rule} fixes the relative orientations of the three classical spins on a triangle at 120$^{\circ }$ from each other in a plane (e.g., see Fig.~\ref{fig:120neel}). But this triangle rule does not fix the relative orientation of the plane of one triad with respect to the planes of the triads on neighboring hyperkagome triangles. These degrees of freedom result in a continuous or local degeneracy of the ground states, among which some allowed noncoplanar magnetic orders will be energetically favorable when $J_{tri}$ and $J_{chain}$ are finite. 
A spin state satisfying the hyperkagome triangle rule is not translational invariant in general. Moreover, we will demonstrate below that the continuous degeneracy exists even though the lattice translational symmetry is respected, which has already reduced lots of degeneracy.

First, let us consider three classical spins on a triangle as shown in Fig.~\ref{fig:120neel}, where
$\bm{\mbox{A}}=\bm{\hat{x}}$, $\bm{\mbox{B}} = -\frac{1}{2}\bm{\hat{x}}+\frac{\sqrt{3}}{2}\bm{\hat{y}}$, $\bm{\mbox{C}} = -\frac{1}{2}\bm{\hat{x}}-\frac{\sqrt{3}}{2}\bm{\hat{y}}$ are three coplanar unit vectors. Thus, a classical {\it coplanar} ordered ground state can be given by the following spin configurations,
\begin{equation}\label{eq:config1}
\begin{array}{lll}
\bm{S}_{1,6,9,10}=\bm{\mbox{A}},\quad & \bm{S}_{2,5,7,11}=\bm{\mbox{B}},\quad & \bm{S}_{3,4,8,12}=\bm{\mbox{C}}.\end{array}
\end{equation}
Here we consider lattice translationally invariant states, so that the unit cell index $i$ is neglected temporarily and  only the sublattice index $\mu$ is kept.

Second, we twist some spins to obtain noncoplanar states continuously. One way to do it is to twist fours spins $\bm{S}_{4}$, $\bm{S}_{5}$, $\bm{S}_{11}$, and $\bm{S}_{12}$ out of the plane of the triad $(\bm{\mbox{A}},\bm{\mbox{B}},\bm{\mbox{C}})$, while leave other spins unchanged. The twist can be done by a rotation $R^{\theta}_{1}$ around the $\hat{x}$ axis (note that $\bm{\mbox{A}} = \bm{\hat{x}}$),
\begin{equation}
R^{\theta}_{1}=\left(
\begin{array}{ccc}
1    & 0             & 0              \\
0    & \cos\theta    & -\sin\theta    \\
0    & \sin\theta    & \cos\theta     \\
\end{array}\right),
\end{equation}
where $\theta\in[-\pi,\pi]$ is the rotation angle. 
Introducing two new unit vectors $\tilde{\bm{\mbox{B}}}=R^{\theta}_{1}\bm{\mbox{B}}$ and $\tilde{\bm{\mbox{C}}}=R^{\theta}_{1}\bm{\mbox{C}}$, we are able to construct the twisted spin configurations as follows:
\begin{equation}\label{eq:config1a}
\begin{array}{rrr}
\bm{S}_{1,6,9,10}=\bm{\mbox{A}},     & \bm{S}_{2,7}=\bm{\mbox{B}},          & \bm{S}_{3,8}=\bm{\mbox{C}}, \\
\bm{S}_{5,11}=\tilde{\bm{\mbox{B}}}, & \bm{S}_{4,12}=\tilde{\bm{\mbox{C}}}. &
\end{array}
\end{equation}
The spin configuration given by Eq.~\eqref{eq:config1a} gives rise to a {\it noncoplanar} magnetic order for arbitrary rotation angle $\theta$, resulting in a continuous degeneracy of the ground states.

It is worth noting that the continuous degeneracy exists even if we break the lattice translational symmetry or restrict ourselves to coplanar spin states. This huge ground-state degeneracy gives rise to finite residual entropy density and remains the classical system disordered down to the lowest temperature.

\subsection{$J_{tri}>0$ and/or $J_{chain}> 0$}
Now we turn on the other two AFM couplings $J_{tri}$ and $J_{chain}$, which will give rise to other energetically favorable spin configurations. The interplay between these spin configurations and the hyperkagome triangle rule will lead to some classical ordered states. We shall analyze possible ground states in this subsection and explore the whole phase diagram by classical Monte Carlo simulation in the next subsection.

{\em $J_{tri}$ coupling.---}  We first consider the case of $J_{tri}> 0$ and $J_{chain}=0$. The $J_{tri}$ bonds emerge from four types of isolated triangles [see Fig.~\ref{fig:triangle_chain}(a)]. Thus another triangle rule will be imposed on such isolated triangles. We will call the new one an ``{\it isolated triangle rule}" to distinguish it from the ``{\it hyperkagome triangle rule}." The isolated triangle rule also fixes the three classical spins within a plane and the angle between two classical spins at 120$^{\circ}$.

The coexistence of both triangle rules will reduce the degeneracy of classical ground states largely and give rise to limited number of classical ground states apart from a global spin rotation.
The spin configuration given by Eq.~\eqref{eq:config1} satisfies both triangle rules and is lattice translationally invariant. In other words, it is a state with lattice wave vector $\bm{Q}=0$. We call such an ordered state as a  ``coplanar $\bm{Q}=0$ state." Note that it is the only $\bm{Q}=0$ spin configuration satisfying both triangle rules apart from a global spin rotation.

Moreover, we find that other possible spin configurations satisfying both triangle rules are with finite lattice wave vectors $\bm{Q}=\frac{2}{3}(\pm\pi,\pm\pi,\pm\pi)$. One of these states is given by $\bm{Q}=\frac{2}{3}(\pi,\pi,\pi)$ and the following spin vectors,
\begin{equation}\label{eq:config2}
\bm{S}_{\bm{r}\mu}=\bm{\hat{x}}\cos(\bm{Q}\cdot\bm{r}+\phi_{\mu})+\bm{\hat{y}}\sin(\bm{Q}\cdot\bm{r}+\phi_{\mu}),
\end{equation}
where $\bm{r}$ denotes a unit cell, $\phi_{1,3,6,10}=0$, $\phi_{2,7,9,11}=4\pi/3$, and $\phi_{4,5,8,12}=2\pi/3$. Then $C_2$ and $C_3$ rotations will give rise to other $\bm{Q}$'s and associated $\phi_{\mu}$'s that satisfy both triangle rules. Note that the state given in Eq.~\eqref{eq:config2} is a coplanar state too. Therefore, a classical ground state is a coplanar state with a wave vector $\bm{Q}=0$ or $\bm{Q}=\frac{2}{3}(\pm\pi,\pm\pi,\pm\pi)$, when $J_{tri}>0$ and $J_{chain}=0$.

{\em $J_{chain}$ coupling.---} The AFM coupling $J_{chain}$ prefers to antiparallel spin aligning along the uniform chains [defined by Eq.~\eqref{eq:chain} and also see Fig.~\ref{fig:triangle_chain}(b)]. Note that such antiparallel spin alignment gives rise to a $\bm{Q}=0$ state, since each unit cell consists of two sites in a uniform chain. We are able to construct such an antiparallel spin configuration satisfying the hyperkagome triangle rule in terms of six unit vectors $\bm{v}_{1}, \bm{v}_{2}, \cdots, \bm{v}_{6}$. To do it, we twist some spins in the state defined in Eq.~\eqref{eq:config1}.
Let $\bm{v}_{1}=\bm{\mbox{A}}$, $\bm{v}_{2}=\bm{\mbox{B}}$ and $\bm{v}_{3}=\bm{\mbox{C}}$, and introduce three new unit vectors, $\bm{v}_{4}$, $\bm{v}_{5}$, and $\bm{v}_{6}$. Then the $\bm{Q}=0$ state is given by
\begin{equation}\label{eq:config3}
\begin{array}{ll}
\bm{S}_{1}=-\bm{S}_{9}=\bm{v}_{1},  & \bm{S}_{2}=-\bm{S}_{11}=\bm{v}_{2}, \\
\bm{S}_{3}=-\bm{S}_{4}=\bm{v}_{3},  & \bm{S}_{5}=-\bm{S}_{7}=\bm{v}_{4},  \\
\bm{S}_{6}=-\bm{S}_{10}=\bm{v}_{5}, & \bm{S}_{8}=-\bm{S}_{12}=\bm{v}_{6},
\end{array}
\end{equation}
where
\begin{equation}
\begin{array}{ll}
\bm{v}_{1}=\left(1,0,0\right), & \bm{v}_{2}=\left(-\frac{1}{2},\frac{\sqrt{3}}{2},0\right), \\\bm{v}_{3}=\left(-\frac{1}{2},-\frac{\sqrt{3}}{2},0\right),&
\bm{v}_{4}=\left(-\frac{1}{2},-\frac{\sqrt{3}}{6},\frac{\sqrt{6}}{3}\right),\\\bm{v}_{5}=\left(0,-\frac{\sqrt{3}}{3},-\frac{\sqrt{6}}{3}\right),&\bm{v}_{6}=\left(\frac{1}{2},-\frac{\sqrt{3}}{6},\frac{\sqrt{6}}{3}\right).
\end{array}
\end{equation}
The $\bm{Q}=0$ state defined in Eq.~\eqref{eq:config3} is a noncoplanar state and we shall call it a ``{\em noncoplanar $\bm{Q}=0$ state}" in order to distinguish it from the previous {\em coplanar $\bm{Q}=0$ state} defined in Eq.~\eqref{eq:config1}.

{\em $J_{tri}$ coexists with $J_{chain}$.---} In general, nonvanishing AFM coupling $J_{tri}$ and $J_{chain}$ will lift the degeneracy and lead to magnetically ordered ground states. As we will discuss below, the hyperkagome triangle rule will be violated as long as both $J_{tri}$ and $J_{chain}$ are nonzero and positive, although it can be satisfied when either $J_{tri}=0$ or $J_{chain}=0$. 

\begin{figure}[tbp]
	\centering
	\includegraphics[width=8.4cm]{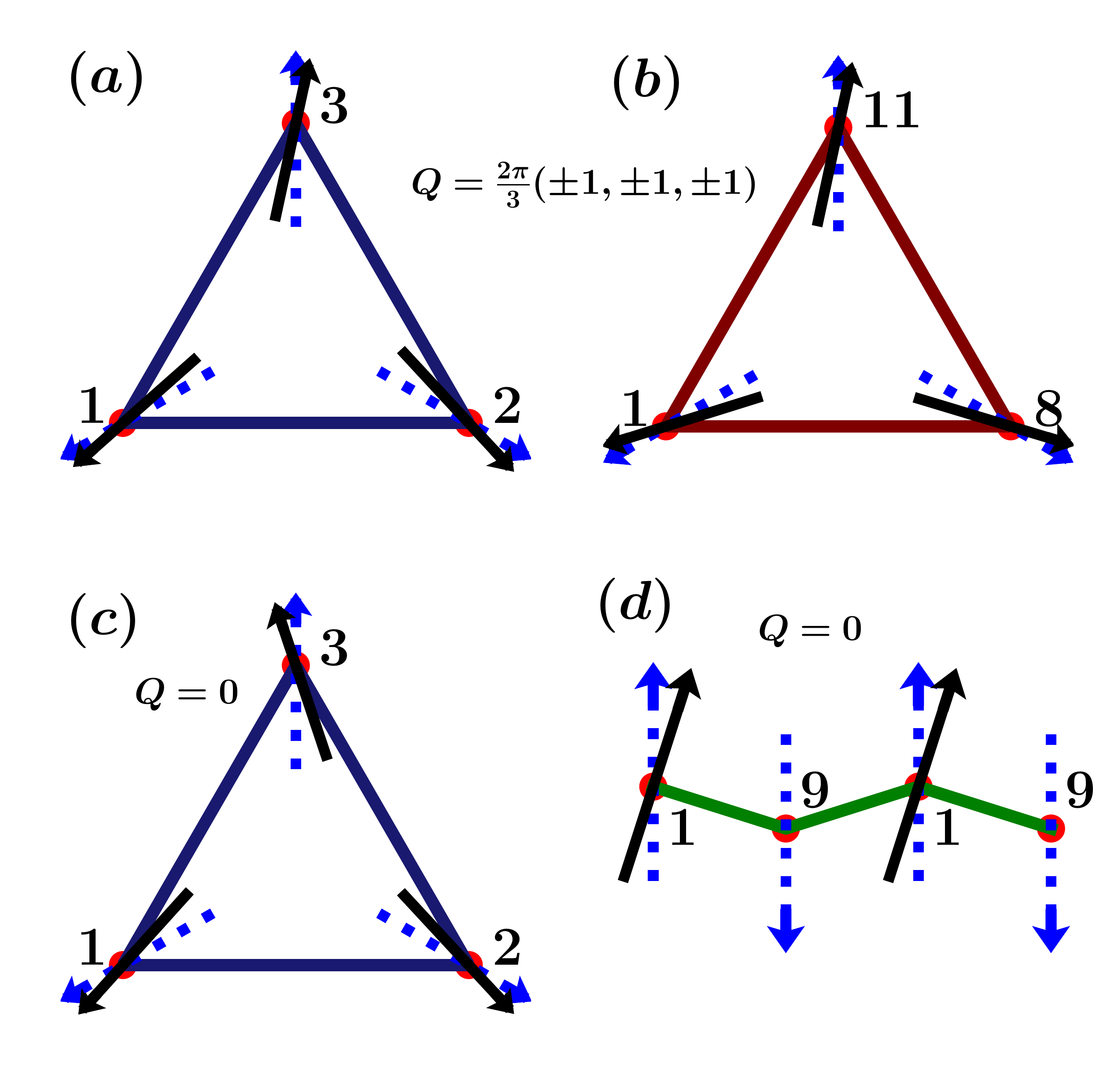}
	\caption{ Spin structures distorted from coplanar states, where solid arrows represent the spins in a coplanar state and dotted arrows represent distorted spins. $\bm{Q}=\frac{2}{3}(\pm\pi,\pm\pi,\pm\pi)$ states: typical spin configurations in (a) hyperkagome triangles and (b) isolated triangles. $\bm{Q}=0$ states: typical spin configurations (c) in hyperkagome triangles and (d) along uniform chains. The distorted spin vectors are no longer coplanar. 
	} 
	\label{fig:distort}
\end{figure}

First, we would like to consider the situation when $J_{tri}\gg J_{chain}$. When $J_{chain}=0$, the $\bm{Q}=0$ coplanar state given by Eq.~\eqref{eq:config1} and the $\bm{Q}=\frac{2}{3}(\pm\pi,\pm\pi,\pm\pi)$ coplanar states given by Eq.~\eqref{eq:config2} are degenerate in energy. However, an infinitesimal AFM $J_{chain}$ will lift such degeneracy and the $\bm{Q}=\frac{2}{3}(\pm\pi,\pm\pi,\pm\pi)$ state will be energetically favorable. Thus, in the presence of a small but finite $J_{chain}>0$, the ground state will be distorted from a coplanar $\bm{Q}=\frac{2}{3}(\pm\pi,\pm\pi,\pm\pi)$ state to a noncoplanar $\bm{Q}=\frac{2}{3}(\pm\pi,\pm\pi,\pm\pi)$ state. Such a noncoplanar state violates both the hyperkagome triangle rule and the isolated triangle rule, and the typical distorted spin configurations on these triangles are illustrated in Figs.~\ref{fig:distort}(a) and \ref{fig:distort}(b). 

When $J_{chain}\gg J_{tri}$, the perturbation of $J_{tri}$ will also violate the hyperkagome triangle rule and destroy the antiparallel spin alignment as shown in Figs.~\ref{fig:distort}(c) and \ref{fig:distort}(d). Although antiparallel spin alignment is destroyed along a uniform chain, the lattice translational symmetries are still respected; namely, it will become a noncoplanar $\bm{Q}=0$ state. The spins on each hyperkagome triangle are slightly twisted from the $120^{\circ}$ configuration and violate the hyperkagome triangle rule. All these results have been verified by classical Monte Carlo simulations.

{\em Incommensurate magnetic orders.---} The situation will become complicated when $J_{tri}$ and $J_{chain}$ are comparable with each other. It is easy to verify that it is impossible to find a classical spin configuration with antiparallel spin alignment along a uniform chain that satisfies both the hyperkagome triangle rule and the isolated triangle rule simultaneously. Moreover, the competition between $\bm{Q}=\frac{2}{3}(\pm\pi,\pm\pi,\pm\pi)$ states [see Figs.~\ref{fig:distort}(a) and (b)] and noncoplanar $\bm{Q}=0$ states [see Figs.~\ref{fig:distort}(c) and (d)] will lead to incommensurate magnetic orders when both $J_{tri}$ and $J_{chain}$ are sizable. To examine this, we shall carry out classical Monte Carlo simulation to study the $J_{1}$-$J_{2}$-$J_{3}$ Heisenberg model and explore the phase diagram.

\subsection{Monte Carlo simulation and phase diagram}

Classical Monte Carlo simulations, based on the standard heatbath method, are performed to study the $J_{1}$-$J_{2}$-$J_{3}$ Heisenberg model defined in Eq.~\eqref{eq:HJ123} on $L\times L\times L\times 12$ hyperkagome lattices. The periodic boundary condition is imposed in the calculations, and the system size is up to $L=24$. The system is gradually cooled down from high temperatures to low temperatures in order to find out the ground state. A run at each temperature consists of 10$^{6}$ Monte Carlo steps and one step consists of $12L^{3}$ spin-flip processes to avoid the autocorrelation.

\begin{figure}[tbp]
	\centering
	\includegraphics[width=8.4cm]{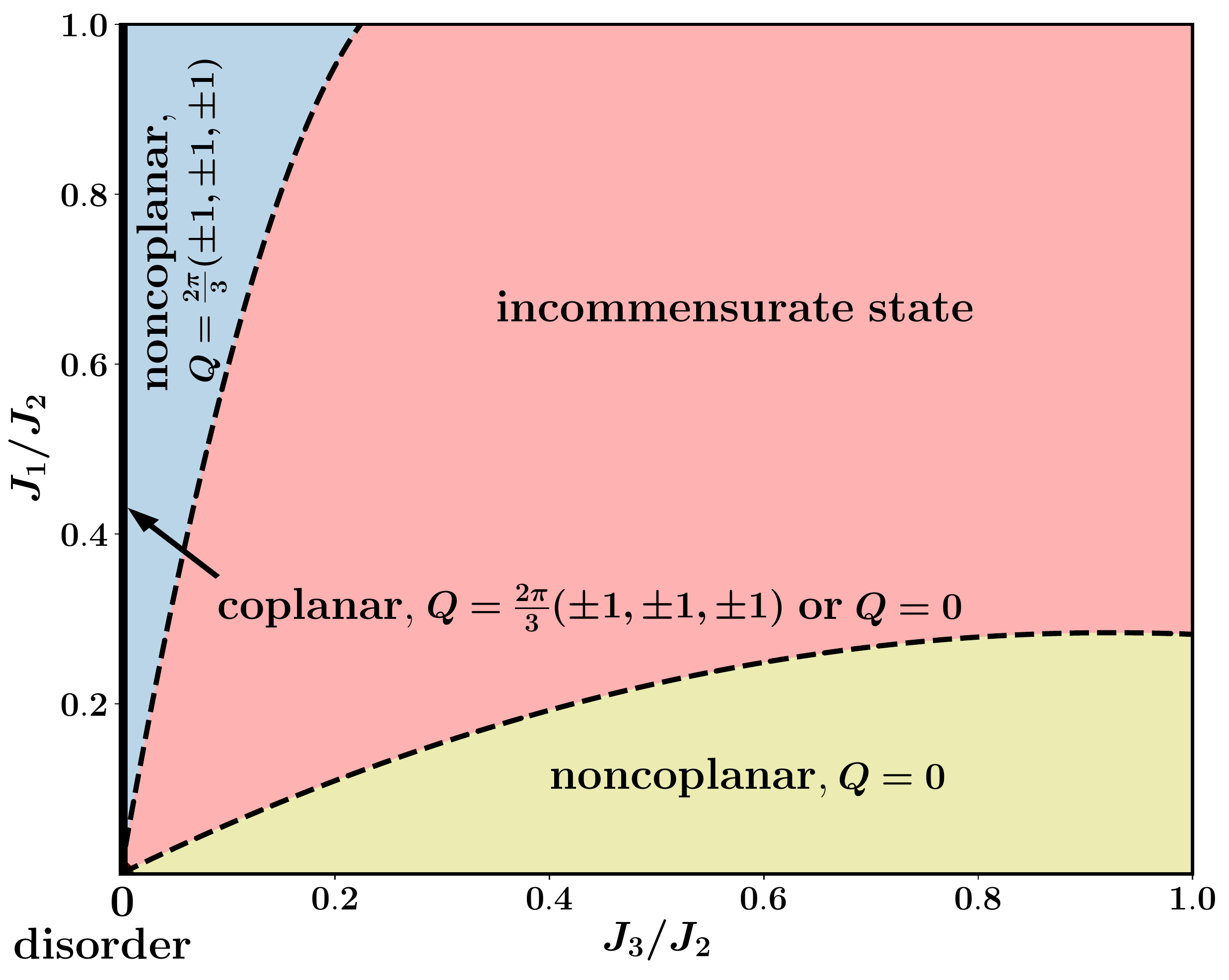}
	\caption{ Sketched ground-state phase diagram for the classical $J_{1}$-$J_{2}$-$J_{3}$ Heisenberg model on a hyperkagome lattice, where $J_{1}=J_{tri}$, $J_{2}=J_{hyper}$, and $J_{3}=J_{chain}$.  There exist five phases: (1) Disordered state in the hyperkagome limit $J_{1}=J_{3}=0$. (2) Along the axis $J_3=0$ and $J_1>0$, the ground state is the coplanar $\bm{Q}=0$ state or the coplanar $\bm{Q}=\frac{2\pi}{3}(\pm 1,\pm 1,\pm 1)$ state, which are degenerate in energy. (3) The noncoplanar $\bm{Q}=\frac{2\pi}{3}(\pm 1,\pm 1,\pm 1)$ states appear on the top left corner, where $J_1\gg J_3$. (4) The noncoplanar $\bm{Q}=0$ state appears in the bottom right corner, where $J_1\ll J_3$. (5) Magnetically ordered states with incommensurate wave vectors $\bm{Q}$ in the middle area, where $J_1$ and $J_3$ are comparable with each other. The two dashed lines indicate the crossovers from incommensurate states to the noncoplanar uniform ($\bm{Q}=0$) state and to noncoplanar $\bm{Q}=\frac{2\pi}{3}(\pm 1,\pm 1,\pm 1)$ states. }
	\label{fig:cphase}
\end{figure}

{\em Phase diagram.---} A sketched phase diagram of ground state is plotted versus $J_{1}/J_{2}$ and $J_{3}/J_{2}$ in Fig.~\ref{fig:cphase}, where $J_{1}=J_{tri}$, $J_{2}=J_{hyper}$, and $J_{3}=J_{chain}$. With the help of Monte Carlo simulations, we find five phases totally: (1) In the hyperkagome limit, $J_{1}=J_{3}=0$, the ground state is magnetically disordered because of the continuous degeneracy and huge residual entropy, which is reminiscent of the AFM Ising model on a triangular lattice~\cite{wannier}. (2) Along the axis $J_{1}=0$, there exist two magnetically ordered ground states. One is the coplanar $\bm{Q}=0$ state given by Eq.~\eqref{eq:config1}, and the other is the coplanar $\bm{Q}=\frac{2\pi}{3}(\pm 1,\pm 1,\pm 1)$ states defined in Eq.~\eqref{eq:config2}. These two ground states are degenerate in energy since both the hyperkagome triangle rule and the isolated triangle rule are satisfied. (3) On the top left corner of the phase diagram where both $J_{1}$ and $J_{3}$ are nonzero and $J_{1}\gg J_{3}$, the ground states are noncoplanar $\bm{Q}=\frac{2\pi}{3}(\pm 1,\pm 1,\pm 1)$ states, which violate both triangle rules [see Figs.~\ref{fig:distort}(a) and (b)]. (4) On the bottom right corner of the phase diagram where $J_{3}\gg J_{1}$, the ground states are noncoplanar $\bm{Q}=0$ states. Along the axis $J_{1}=0$, the ground state is the noncoplanar $\bm{Q}=0$ state given in Eq.~\eqref{eq:config3}, which satisfies the hyperkagome triangle rule and spins are antiparallel aligned along a uniform chain. Away from the axis $J_{1}=0$, the ground state is still a noncoplanar $\bm{Q}=0$ state, but the spin configuration is distorted from that given in Eq.~\eqref{eq:config3}, and the hyperkagome triangle rule is violated and spins are no longer antiparallel aligned along a uniform chain [see Figs.~\ref{fig:distort}(c) and (d)]. (5) In the middle area where $J_1$ and $J_3$ are comparable with each other, the ground state is a magnetically ordered state with an incommensurate wave vector $\bm{Q}$, as we will discuss in detail below.

\begin{table}[tbp]
\caption{
The fractional number $m_p/L_p$ giving rise to the ground state wave vector $\bm{Q}=2\pi m_p/L_p(\pm 1,\pm 1,\pm 1)$. In the Monte Carlo calculations, we use hyperkagome lattices with size $L=1,\cdots,24$. Here we set $J_2=1$.
The $m_p/L_p$ with a footnote mark means that the corresponding fractional number in the footnote will give rise to an approximately degenerate ground state, namely, the two ground state energies are numerically indistinguishable.}\label{tab:incom}
\renewcommand\arraystretch{1.6} 		
\setlength\tabcolsep{0.13cm}
\begin{tabular}{|c|c|c|c|c|c|c|c|c|c|c|}
\hline
\diagbox{$J_{1}$}{$J_{3}$} & $0.1$ & $0.2$ & $0.3$ & $0.4$ & $0.5$ & $0.6$ & $0.7$ & $0.8$ & $0.9$ & $1.0$ \\
\hline
0.2 & \large$\frac{2}{7}$ & \large$\frac{1}{5}$ & \large$\frac{1}{10}$ & \large$\frac{1}{14}$\footnotemark[1] & $0$ & $0$ & $0$ &     $0$ &     $0$ &     $0$\\
\hline
0.4 & \large$\frac{1}{3}$ & \large$\frac{2}{7}$ & \large$\frac{1}{4}$ & \large$\frac{1}{5}$& \large$\frac{1}{7}$ & \large$\frac{1}{7}$\footnotemark[2] & {\large ${\frac{1}{12}}$} & {\large ${\frac{1}{12}}$} &   {\large ${\frac{1}{13}}$\footnotemark[3]}&   {\large ${\frac{1}{13}}$}\\
\hline
0.6 & \large$\frac{1}{3}$ & \large ${\frac{3}{10}}$\footnotemark[4] & \large ${\frac{2}{7}}$ & \large ${\frac{1}{4}}$ & \large ${\frac{2}{9}}$ &  \large ${\frac{1}{6}}$ & \large ${\frac{1}{6}}$ & \large ${\frac{1}{7}}$ &  \multicolumn{2}{c|}{\large$\frac{1}{8}$\footnotemark[5]} \\
\hline
0.8 & \large$\frac{1}{3}$ &  \large$\frac{1}{3}$ & \large ${\frac{3}{10}}$ & \large ${\frac{3}{11}}$ & \large ${\frac{1}{4}}$ &  \large ${\frac{2}{9}}$ &  \large ${\frac{1}{5}}$ & \large $\frac{1}{6}$\footnotemark[6] &  \large${\frac{1}{6}}$ & \large${\frac{1}{6}}$ \\
\hline
\end{tabular}
\footnotemark[1]{\large $\frac{1}{13}$}, 
\footnotemark[2]{\large $\frac{1}{10}$}, 
\footnotemark[3]{\large $\frac{1}{12}$}, 
\footnotemark[4]{\large $\frac{4}{13}$}, 
\footnotemark[5]{\large $\frac{1}{9}$}, 
\footnotemark[6]{\large $\frac{2}{11}$}. 
\end{table}

\begin{table}[tbp]
	\caption{ The ground state energy $\epsilon_L$ for finite $L\times L\times L \times 12$ lattices (up to $L=L_{max}=24$) and corresponding fractional number $m_p/L_p$ defined in Eq.~\eqref{eq:vecQ}. We set $J_{2}=1$ and choose $J_{1}/J_{2}=0.6$. }
\label{tab:mcdata}
	\renewcommand\arraystretch{1.25}
	\setlength\tabcolsep{0.3cm}
	\begin{tabular}{|c|c|c|c|c|c|}
		\hline
		\multicolumn{3}{|c|}{$\ J_{3}/J_{2}=0.8$} & \multicolumn{3}{c|}{$\ J_{3}/J_{2}=0.3$} \\
		\hline
		$L$ & $\epsilon_L$ & $m_p/L_p$ & $L$ & $\epsilon_L$ &  ${m_p}/{L_p}$ \\
		\hline
		6 & -1.634 & 1/6         & 4  & -1.359 & 1/4 \\
		7 & -1.643 & ${1}/{7}$   & 6  & -1.351 & 1/3 \\
		8 & -1.633 & ${1}/{8}$   & 7  & -1.362 & 2/7 \\
		9 & -1.634 & ${1}/{9}$   & 8  & -1.359 & 1/4 \\
		10 & -1.631 & ${1}/{10}$ & 9  & -1.351 & 1/3\\
		12 & -1.633 & ${1}/{6}$  & 11 & -1.354 & 3/11\\
		13 & -1.634 & ${2}/{13}$ & 12 & -1.359 & 1/4\\
		14 & -1.642 & ${1}/{7}$  & 14 & -1.362 & 2/7 \\
		21 & -1.641 & ${1}/{7}$  & 21 & -1.362 & 2/7\\
		22 & -1.633 & ${3}/{22}$ & 22 & -1.353 & 3/11\\
		23 & -1.634 & ${3}/{23}$ & 23 & -1.352 & 6/23\\
		24 & -1.632 & ${1}/{8}$  & 24 & -1.357 & 1/4\\
		\hline
	\end{tabular}
\end{table}

{\em Incommensurability.---} In the middle region of the phase diagram where $J_{1}$ and $J_{3}$ are comparable with each other, we find numeric evidences for incommensurate magnetic ordering. In the Monte Carlo simulations, we set $J_{2}=1$ and change $J_1$ and $J_3$ to calculate ground state energy per site $\epsilon_{L}$, the spin correlation function $$C(\bm{r}) = \frac{1}{12}\sum_{\mu} \langle \bm{S}_{i+\bm{r},\mu}\cdot\bm{S}_{i,\mu} \rangle,$$ and its Fourier transformation $$S(\bm{q})= \sum_{\bm{r}} e^{i\bm{q}\cdot\bm{r}} C(\bm{r})$$ on hyperkagome lattices up to $L=24$. 
In the middle region (see Fig.~\ref{fig:cphase}), the static structure factor $S(\bm{q})$ always exhibits peaks at finite wave vectors $\bm{q}=\bm{Q}$ in the directions of $(\pm 1,\pm 1,\pm 1)$. For a finite $L\times L\times L\times 12$ lattice and all the parameters except $J_1=J_3=0$, we always find that
\begin{equation}\label{eq:vecQ}
\bm{Q}=\frac{2\pi m}{L}(\pm 1,\pm 1,\pm 1)=\frac{2\pi m_p}{L_p}(\pm 1,\pm 1,\pm 1),
\end{equation}
where $0\le m < L$, $L_p$ is an integer factor of $L$, and two integers $m_p$ and $L_p$ are coprime, i.e., $(m_p,L_p)=1$. So that the period of the magnetic ordering is $L_p$ in a $L\times L\times L\times 12$ hyperkagome lattice.

For a given pair of $(J_{1},J_{3})$, we change the lattice size up to $L=L_{max}$ to find out the lowest ground-state energy $\epsilon_{min}$ among all these $L\times L\times L \times 12$ lattices, namely,
\begin{equation}\label{eq:Lmin}
\epsilon_{min}=\text{min}\{\epsilon_{L}|L=1,\cdots,L_{max}\},
\end{equation}
and the corresponding fractional number $\frac{m_p}{L_p}$ giving rise to the wave vector $\bm{Q}$ in Eq.~\eqref{eq:vecQ}, where $L_{max}=24$ is the upper bound of the linear size $L$ that we used in Monte Carlo calculations. 
The values of fractional number $\frac{m_p}{L_p}$ are listed in Table~\ref{tab:incom}. 
As examples, we consider two points  on the phase diagram, say, $(J_{1}/J_{2},J_{3}/J_{2})=(0.6,0.3)$ and $(J_{1}/J_{2},J_{3}/J_{2})=(0.6,0.8)$, and list the ground-state energy $\epsilon_{L}$ up to $L=24$ in Table~\ref{tab:mcdata}. For $L_{max}=24$, we have $\epsilon_{min}=-1.362J_{2}$ and $m_{p}/L_{p}=2/7$ at $(J_{1}/J_{2},J_{3}/J_{2})=(0.6,0.3)$, while $\epsilon_{min}=-1.643J_{2}$ and $m_{p}/L_{p}=1/7$ at $(J_{1}/J_{2},J_{3}/J_{2})=(0.6,0.8)$.
It is also found that the rational number $m_p/L_p$ changes from point to point within this phase, so that we expect that $m_p/L_p$ will approach an irrational number in general in the thermodynamic limit $L_{max}\to \infty$. 
It means that the wave vector $\bm{Q}$ varies with $(J_1,J_3)$ in the middle area and is incommensurable generally.

\section{Schwinger Boson Theory}\label{sec:sboson}

In the remaining part of this paper, we will study the effect of quantum fluctuations with the help of Schwinger bosons. The Schwinger boson theory may give rise to a gapful spin liquid state with low-lying bosonic spinon excitations, or a magnetically ordered state when bosonic spinons are condensed at some gapless modes. In such an ordered state, a gapless magnon excitation is made of two gapless bosonic spinons. In this section, we shall set up Schwinger boson theory in the large-$N$ formulation and analyze possible fractionalized states using projective symmetry group (PSG)~\cite{PSG2}. 

\subsection{Schwinger boson representation, large-$N$ formulation, and mean-field theory}

In the quantum theory, the model Hamiltonian is still given by Eq.~\eqref{eq:HJ123}, but all the classical spin vectors $\bm{S}_{i\mu}$ will be replaced by spin operators $\hat{\bm{S}}_{i\mu}$. We introduce two species of Schwinger bosons $b_{i\mu\alpha}$ ~\cite{largeN1,largeN2,largeN3,largeN4} ($\alpha=\uparrow,\downarrow$) for a quantum spin $S$ as follows:
\begin{subequations}
	\begin{equation}
	\hat{\bm{S}}_{i\mu}=\frac{1}{2}\sum_{\alpha,\beta} b_{i\mu\alpha}^{\dagger} \bm{\sigma}_{\alpha\beta}  b_{i\mu\beta},
	\end{equation}
	with the constraint imposed at each site $i\mu$,
	\begin{align}
	&\hat{n}_{i\mu}=\sum_{\alpha} b_{i\mu\alpha}^{\dagger}b_{i\mu\alpha}=n_{b},\label{eq:cstrnt}
	\end{align}
\end{subequations}
where $n_{b}$=$2S$ is the number of Schwinger bosons per site, $\alpha,\beta = |\uparrow\rangle, |\downarrow\rangle$ and $\bm{\sigma}$ are three Pauli matrices. Then the spin exchange interactions in Eq.~\eqref{eq:HJ123} can be rewritten in terms of Schwinger bosons as follows:
\begin{subequations}
	\begin{equation}
	\hat{\bm{S}}_{i\mu}\cdot\hat{\bm{S}}_{j\nu} = -\frac{1}{2}\hat{A}^{\dagger}_{i\mu{}j\nu}\hat{A}_{i\mu{}j\nu} + \frac{1}{4}\hat{n}_{i\mu{}}\hat{n}_{j\nu},
	\end{equation}
	where valence-bond operators $\hat{A}_{i\mu{}j\nu}$ are defined as
	\begin{equation}
	\hat{A}_{i\mu{}j\nu} = b_{i\mu{}\uparrow}b_{j\nu\downarrow} -  b_{i\mu{}\downarrow}b_{j\nu\uparrow}.
	\end{equation}
\end{subequations}
Note that $\hat{A}_{i\mu{}j\nu}=-\hat{A}_{j\nu{}i\mu}$ for a bosonic representation.

The SU(2) valence-bond operator $\hat{A}_{i\mu{}j\nu}$ can be generalized into Sp(N) case by introducing $2N$ flavor Schwinger boson $b_{i\mu\alpha}$ and becomes
\begin{equation}
\hat{A}_{i\mu{}j\nu}=\sum_{\alpha\beta}\mathcal{J}_{\alpha\beta}b_{i\mu\alpha}b_{j\mu\beta},
\end{equation} 
where $\mathcal{J}$ is a $2N\times{}2N$ matrix and reads
\begin{equation}
\mathcal{J}=\left(\begin{array}{cccccc}
& 1 & & & & \\
-1 &  & \\
& & & 1 & & \\
& &-1 &  & & \\
& & & & &  \ddots\\
\end{array}\right),
\end{equation} 
and the index $\alpha$=$1$, \dots, $2N$ is transformed under the symplectic group Sp(N)~\cite{largeN1,largeN2,largeN3,largeN4}. For a realistic material, we shall focus on its physical realization $N=1$, when Sp(1) is isomorphic to SU(2). However, generic situations with $N>1$ are also of great interest, since the mean-field theory will be exact in the large-$N$ limit. It is convenient to introduce the following parameter,
\begin{equation}\label{eq:kappa}
\kappa=\frac{n_{b}}{N}. 
\end{equation}
Thus, the large-$N$ limit is taken with fixed $\kappa$. In this paper, we consider $\kappa$ as a positive and continuous parameter, which will be $\kappa=1$ when $S=1/2$ and $N=1$. For our $J_1$-$J_2$-$J_3$ model, it turns out that large $\kappa$ will lead to a magnetically ordered state while small $\kappa$ will give rise to quantum spin liquid states.

The Schwinger boson mean-field theory, which will become exact in the limit $N\to \infty$, can be formulated in accordance with the following Hamiltonian,
\begin{equation}\label{eq:HMF}
\begin{split}
H_{MF}=&\sum_{a=1}^{3}\sum_{\langle{}i\mu,j\nu\rangle_{a}}\frac{J_{a}}{2}\Big{[}-(A_{i\mu{}j\nu}\hat{A}^{\dagger}_{i\mu{}j\nu} + \text{H.c.}) \\& + |A_{i\mu{}j\nu}|^{2} + \frac{1}{4} \ \Big{]} + \lambda\sum_{i\mu\alpha}(b_{i\mu{}\alpha}^{\dagger}b_{i\mu\alpha} - \kappa N),
\end{split}
\end{equation}
where $A_{i\mu{}j\nu}=-A_{j\nu{}i\mu}$ are complex numbers, and the Lagrange multiplier $\lambda$ is a real number. 
The set of parameters, $\{A_{i\mu{}j\nu},\lambda\}$, is called {\em mean-field ansatz}, which can be determined by the following self-consistent equations,
\begin{subequations}\label{eq:self}
	\begin{equation}\label{eq:Aij}
	A_{i\mu{}j\nu} = \langle{}\hat{A}_{i\mu{}j\nu}\rangle,
	\end{equation}
	and
	\begin{equation}
	\frac{\langle{}\hat{n}_{i\mu{}}\rangle}{N}=\kappa,
	\end{equation}
\end{subequations}
where $\langle{}\cdots\rangle$ means that the expectation is evaluated in the ground state of the mean-field Hamiltonian $H_{MF}$.
Note that for a given set of mean-field ansatz, the mean-field Hamiltonian given in Eq.~\eqref{eq:HMF} can be solved with the help of Bogoliubov transformation and singular value decomposition. The details of the self-consistent calculation can be found in Appendix \ref{ap:selfcon}.

\subsection{Gauge structure and PSG}

In the Schwinger boson mean-field theory, the particle number constraint in Eq.~\eqref{eq:cstrnt} is implemented on average. A physical spin state $|\Psi_{Spin}\rangle$ can be constructed from the mean-field ground state $|\Psi_{MF}(A_{i\mu{}j\nu},\lambda)\rangle$ via a Gutzwiller projection denoted by $P_G$,
\begin{equation}
|\Psi_{Spin}\rangle=P_G|\Psi_{MF}(A_{i\mu{}j\nu},\lambda)\rangle.
\end{equation}
The Gutzwiller projection $P_G$ keeps the wavefunction components with $n_b=2S$ bosons at each site only and removes other components from $|\Psi_{MF}(A_{i\mu{}j\nu},\lambda)\rangle$.
It means that the particle number constraint is satisfied in the projected state strictly.

{\em $U(1)$ gauge structure.---} There exists $U(1)$ gauge redundancy in the Schwinger boson representation of spins. Namely, the physical spin operators $\bm{S}_{i\mu}$ and thereby other physical observables will not change under the local $U(1)$ gauge transformation,
\begin{subequations}\label{eq:gaugeU1}
	\begin{equation}\label{eq:gaugeb}
	b_{i\mu\alpha}\rightarrow{}e^{i\phi_{i\mu}}b_{i\mu\alpha}.
	\end{equation}
	On the other hand, the mean-field Hamiltonian is invariant under the $U(1)$ gauge transformation defined in Eq.~\eqref{eq:gaugeb} and the compensated gauge transformation for $A_{i\mu{}j\nu}$,
	\begin{equation}\label{eq:gaugeAij}
	A_{i\mu{}j\nu}\rightarrow{}e^{-i\phi_{i\mu}-i\phi_{j\nu}}A_{i\mu{}j\nu}.
	\end{equation}
\end{subequations}
Hence two mean-field ansatzes $\{A_{i\mu{}j\nu},\lambda\}$ and $\{\tilde{A}_{i\mu{}j\nu},\lambda\}$ related by a gauge transformation will give rise to the same physical spin state $|\Psi_{Spin}\rangle$, although corresponding mean-field ground states $|\Psi_{MF}(A_{i\mu{}j\nu},\lambda)\rangle$ can differ from each other.

{\em Invariant gauge group} (IGG).--- The gauge transformations that do not change the mean-field ansatz $\{A_{i\mu{}j\nu},\lambda\}$ constitute a subgroup of the original $U(1)$ gauge group called an {\em invariant gauge group}.
For the Schwinger boson mean-field Hamiltonian given in Eq.~\eqref{eq:HMF}, the IGG should be a $Z_2$ group with elements $\{1,-1\}$ as long as $\{A_{i\mu{}j\nu}\}$ does not vanish, since the pairing of two bosonic spinons always breaks $U(1)$ to $Z_2$.

{\em Projective symmetry group} (PSG).--- In analogy to Landau's symmetry classification for classical orders, it was proposed by Wen~\cite{PSG2} that the symmetry of the mean-field ansatz $\{A_{i\mu{}j\nu},\lambda\}$ is a universal property and can be used to characterize quantum orders in quantum spin liquid states. 
The mathematical tool to characterize these quantum orders is the {\em projective symmetry group}.
An element of a PSG is a combined operation consisting of a symmetry transformation $U$ followed by a local gauge transformation $G_{U}(i\mu)$. 
The PSG of a given mean-field ansatz consists of all combined operations that leave the ansatz unchanged, i.e.,
\begin{equation}\label{eq:PSGdef}
\mbox{PSG}\equiv \{G_{U}|G_{U}U(A_{i\mu{}j\nu})=A_{i\mu{}j\nu}, G_{U}(i\mu)\in U(1)\},
\end{equation}
where $U(A_{i\mu{}j\nu})=\tilde{A}_{i\mu{}j\nu}\equiv A_{U(i\mu){}U(j\nu)}$, $G_{U}U(A_{i\mu{}j\nu})\equiv G_{U}(i\mu)\tilde{A}_{i\mu{}j\nu}G_{U}(j\nu)$, $U$ is an element of the symmetry group which generates the symmetry transformation, and $G_{U}$ is the associated gauge transformation. It is worth noting that IGG is a special subgroup of PSG. 

In this paper, we shall classify the bosonic mean-field ansatz $\{\tilde{A}_{i\mu{}j\nu},\lambda\}$ on hyperkagome lattices in accordance with their PSGs. 

\subsection{PSGs for bosonic states with $P4_{1}32$ symmetry}

We shall find out all the PSGs for bosonic states with $P4_{1}32$ symmetry. As mentioned, the Schwinger boson mean-field Hamiltonian defined in Eq.~\eqref{eq:HMF} always gives rise to a $Z_2$ state, so that we only consider $Z_2$ PSGs in this paper. Following Wen's strategy~\cite{PSG2}, we shall derive algebraic constraints for the PSG elements at first, which allows us to construct bosonic mean-field ansatzes according to obtained algebraic PSGs. The idea is that the algebraic relations among physical symmetry operations will impose algebraic constraints on the structure of PSGs, which can be used to derive all the possible PSGs instead of referring to a specific mean-field ansatz.

We consider the PSG classification for bosonic states with $P4_{1}32$ space group. Note that a similar PSG classification for fermionic states has been done by Huang {\it et al.} in Ref.~[\onlinecite{huang17}], where time-reversal symmetry is considered as well. In this paper, we will not involve time-reversal symmetry and consider space group symmetry $P4_{1}32$ only, such that time-reversal symmetry-breaking states are allowed indeed. 

As discussed in previous subsection, an operation in PSG is implemented via combining a physical symmetry operation with a gauge transformation. For $P4_{1}32$ space group symmetry, it is sufficient to consider the six generators among all the group elements, namely, three lattice translations $T_{1,2,3}$ defined in Eq.~\eqref{eq:T123}, screw rotation defined in Eq.~\eqref{eq:S4}, and $C_2$ and $C_3$ rotations defined in Eqs.~\eqref{eq:C2} and \eqref{eq:C3} respectively. Associated gauge transformations $G_{U}$ now are $G_{T_{1,2,3}}$, $G_{S_{4}}$, $G_{C_{2}}$, and $G_{C_{3}}$ respectively.

The space group will impose algebraic constraints on these $G_{U}$'s. For instance, from $$T_{1}^{-1}T_{2}T_{1}T_{2}^{-1}=I$$ we know that the implementation of such operator, i.e., $$(G_{T_{1}}T_{1})^{-1}G_{T_{2}}T_{2}G_{T_{1}}T_{1}(G_{T_{2}}T_{2})^{-1}$$ should be equivalent to $G_{I}I$, which is nothing but an element of the IGG. This leads to an algebraic constraint, $$G_{T_{1}}^{-1}(T_{1}(i))G_{T_{2}}(T_{1}(i))G_{T_{1}}(T_{1}T_{2}^{-1}(i))G_{T_{2}}^{-1}(i)=\pm{}1,$$ where the sublattice index $\mu$ is neglected for brevity because the operations $T_{1,2,3}$ do not couple two sublattices. Other algebraic relations involving screw rotation $S_4$ and $C_2$ and $C_3$ rotations can be found Appendix \ref{ap:PSG}, as well as detailed calculations. The algebra is straightforward and is similar to that used in a fermionic PSG calculation on a  hyperkagome lattice~\cite{huang17}. Additionally, for bosonic states, there exists a convenient gauge choice by which all the gauge rotations $G_{U}(i\mu)$ are independent of the unit cell index $i$. This implies that there exist only A-type ansatzes~\cite{PSG2} for $Z_2$ bosonic states, in which not only the projected spin states but also the ansatzes themselves respect the lattice translational symmetry.

Eventually, we find four different PSGs belonging to two classes as follows:
\begin{subequations}
	\begin{align}
	\begin{split}\label{eq:psg1}
	\mbox{Class 1:}\quad&\\
	&g_{C_{2}} = \eta_{0},\quad g_{S_{4}} = 1,\\
	&g_{C_{3}}(1,4,7,5,8,10)=1, \\&g_{C_{3}}(2,3,6,9,11,12)=\eta_{0},
	\end{split}\\
	\begin{split}\label{eq:psg2}
	\mbox{Class 2:}\quad&\\
	&g_{C_{2}}(1,4,7,9,12,2)  = i\eta_{0}, \\
	&g_{C_{2}}(5,8,10,6,11,3)  = -i\eta_{0},\\
	&g_{C_{3}}(1,4,7) = 1,\quad g_{C_{3}}(5,8,10)=-1,\\ &g_{C_{3}}(2,3,6,9,11,12)=i\eta_{0},\\
	&g_{S_{4}} = 1,
	\end{split}
	\end{align}
\end{subequations}
where $g_{U}(\mu)\equiv{}G_{U}(i=0,\mu)\in Z_2$, and  $\eta_{0}=\pm{}1$ gives rise to four different algebraic PSGs. It will be seen next subsection that $\eta_0=\pm 1$ will give rise to the same type of mean-field ansatzes when we consider the first three NN bonds only, which can be distinguished by gauge-invariant fluxes.

\subsection{From algebraic PSG to mean-field ansatz}
Now we shall discuss possible bosonic mean-field ansatzes allowed by PSGs in Eqs.~\eqref{eq:psg1} and \eqref{eq:psg2}.  A symmetric mean-field ansatz can be generated as follows. Without loss of generality, we begin with a bond $\langle i\mu,j\nu\rangle$ with a nonzero order parameter $A_{i\mu{}j\nu}=-A_{i\mu{}j\nu}=\Delta$, where $\Delta$ is a complex number. Thus the order parameters $A_{i\mu{}j\nu}$ on other bonds can be generated by
\begin{equation}\label{eq:p2a}
A_{U(i\mu)U(j\nu)} = G^{-1}_{U(i\mu)}A_{i\mu{}j\nu}G^{-1}_{U(j\nu)},
\end{equation}
where the operation $U$ runs over all the 24 group elements of P4$_{1}$32.
For the $J_1$-$J_2$-$J_3$ Heisenberg model, we shall restrict ourselves to first, second, and third NN bonds, namely, we consider the mean-field ansatzes with nonzero $A_{i\mu{}j\nu}$ only on these bonds, while letting $A_{i\mu{}j\nu}=0$ on other bonds.

\begin{figure}[tpb]
	\includegraphics[width=8.0cm]{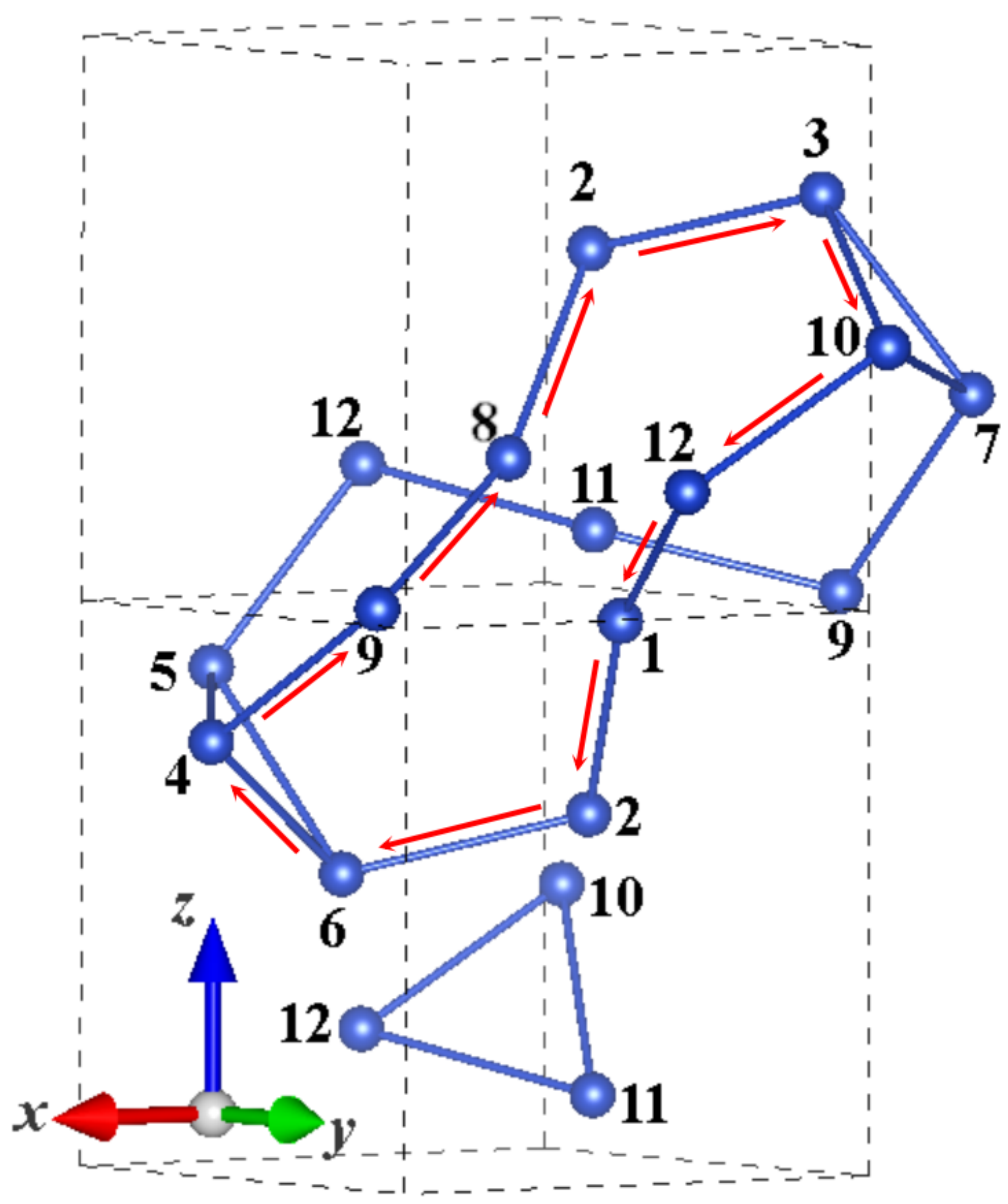}
	\caption{ An example for the elementary 10-site loops that gives rise to the gauge-invariant flux $\Phi=\Phi_{10}$. }\label{fig:bigloops}
\end{figure}

\emph{Gauge-invariant flux.---} With the help of PSGs, we find that there are two types of mean-field ansatzes, which can be distinguished by the gauge-invariant flux~\cite{tchernyshyov06,wang06} $\Phi$ on a closed loop ${\cal C}=\{ i_1\mu_1,i_2\mu_2,\cdots,i_{2k-1}\mu_{2k-1},i_{2k}\mu_{2k}\}$,
\begin{eqnarray}
\Phi&=&\arg[A_{i_1\mu_1{}i_2\mu_2}(-A_{i_2\mu_2{}i_3\mu_3}^{*})A_{i_3\mu_3{}i_4\mu_4}\nonumber\\
&&\times\cdots{}A_{i_{2k-1}\mu_{2k-1}{}i_{2k}\mu_{2k}} (-A_{i_{2k}\mu_{2k}{}i_1\mu_1}^{*})], \label{eq:defPhi}
\end{eqnarray}
where the antisymmetric relation, $A_{i\mu{}j\nu}=-A_{j\nu{}i\mu}$, has been accounted. Note that such a gauge-invariant flux $\Phi$ can be defined only in the loops with even number of bonds [see Eq.~\eqref{eq:gaugeAij} as well], which is very different from usual gauge-invariant SU(2), $U(1)$, or $Z_2$ fluxes defined in fermionic states. It means that such a flux $\Phi$ can not be defined on any triangles. Since the bonds on the uniform chains can not form any closed loop, it is sufficient to consider closed loops on the hyperkagome network only. It is known that all the closed loops on a hyperkagome network can be decomposed into three-site loops (hyperkagome triangles) and elementary 10-site loops. An example of such an elementary 10-site loop $(6\rightarrow4\rightarrow9\rightarrow8\rightarrow2\rightarrow3\rightarrow10\rightarrow12\rightarrow1\rightarrow2\rightarrow6)$ is demonstrated in Fig.~\ref{fig:bigloops}.
Moreover, two hyperkagome triangles and three elementary 10-site loops will form a closed surface~\cite{NaIrO4}. Therefore, we can parametrize these two types of ansatzes by the flux $\Phi\equiv \Phi_{10}$, where $\Phi_{10}$ is the gauge-invariant flux on an elementary 10-site loop.  When $\Phi$=$0$, we call the ansatz a {\em zero-flux state}, while when $\Phi$=$\pi$, we call it a {\em $\pi$-flux state}.

\emph{$\pi$-flux state.---}
The $\pi$-flux state is given by the algebraic PSG in Eq.~\eqref{eq:psg1}. With the help of the symmetric operations $S_{4}$, $C_2$ and $C_3$, we find that $A_{i\mu{}j\nu}$ on the first and the third NN bonds will vanish and the remaining nonzero $A_{i\mu{}j\nu}$ will appear on the second NN (hyperkagome) bonds only (referring to $J_{2}$ couplings). Consider the first NN bond $(i\mu,j\nu)=(1,8)$ and the third NN bond $(1,9)$ (here we omit the unit-cell indices for short without causing confusion), $A_{1,8}$ is transferred as,
\begin{equation*}
A_{1,8}\stackrel{S_{4}}{\longrightarrow}A_{6,7}\stackrel{C_{3}}{\longrightarrow}A_{8,1},
\end{equation*}
under successive transformations $S_4$ and $C_3$, and $A_{1,9}$ is transferred as
\begin{equation*}
A_{1,9}\stackrel{C_{2}}{\longrightarrow}A_{9,1},
\end{equation*}
under $C_2$. Taking into account the antisymmetric relation $A_{i\mu{}j\nu}=-A_{j\nu{}i\mu}$, we obtain $A_{1,8}=A_{1,9}=0$. Symmetry operations of $P4_{1}32$ will generate all the other first and third NN bonds; therefore, we draw the conclusion that $A_{i\mu{}j\nu}=0$ on all the first and third NN bonds, and the mean-field ansatz for a $\pi$-flux state can be written as
\begin{equation}
\begin{split}
\left.\begin{array}{r}
A_{1,2}=A_{1,12}=A_{6,5}=A_{6,8}\\
A_{9,7}=A_{9,4}=A_{10,3}=A_{10,11}
\end{array}\right\rbrace&=\Delta_{2},\\
\left.\begin{array}{r}
A_{2,3}=A_{3,1}=A_{5,4}=A_{4,6}\\
A_{8,2}=A_{2,6}=A_{7,8}=A_{8,9}\\
A_{12,5}=A_{5,1}=A_{3,7}=A_{7,10}\\
A_{4,11}=A_{11,9}=A_{11,12}=A_{12,10}
\end{array}\right\rbrace&=\eta_{0}\Delta_{2},
\end{split}\label{eq:ansatz1}
\end{equation}
where $\eta_{0}=\pm{}1$ and $\Delta_{2}$ is the short-range order parameter on the second NN (hyperkagome) bonds. Moreover, the gauge transformation 
\begin{equation}\label{eq:gauge}
G_{i\mu} = \left\lbrace\begin{array}{ll}
1, & \mbox{when } \mu =1,2,4,7,9,12, \\
-1, & \mbox{when } \mu =3,5,6,8,10,11,\\
\end{array} \right.
\end{equation}
will change the sign of $\eta_{0}$ in Eq.~\eqref{eq:ansatz1} while leaving $\Delta_2$ unchanged, say, $$\eta_0\longrightarrow -\eta_0,\,\, \Delta_2 \longrightarrow \Delta_2,$$ which allows us to consider the ansatz with $\eta_{0}=1$ only. For such a $\pi$-flux state, $\Delta_{2}$ can always be chosen as a real number because of the $U(1)$ gauge redundancy, which indicates that the time-reversal symmetry is respected.

\emph{Zero-flux state.---}
The zero-flux state is given by the algebraic PSG in Eq.~\eqref{eq:psg2}. The mean-field ansatz $A_{i\mu{}j\nu}$ (here we omit indices $i$ and $j$ for short) reads
\begin{subequations}\label{eq:ansatz2}
	\begin{equation}\label{eq:ansatz21}
	\begin{split}
	\left.\begin{array}{r}
	A_{11,1} = A_{1,8} = A_{2,10} = A_{10,4}\\
	A_{5,9} = A_{9,3} = A_{7,6} = A_{6,12}
	\end{array}\right\rbrace&=\Delta_{1},\\
	A_{11,8} = A_{4,2} = A_{3,5} = A_{12,7} &= \eta_{0}\Delta_{1},\\
	\end{split}
	\end{equation}
	on the first NN bonds, 
	\begin{equation}\label{eq:ansatz22}
	\begin{split}
	\left.\begin{array}{r}
	A_{12,1} = A_{1,2} = A_{8,6} = A_{6,5}\\
	A_{4,9} = A_{9,7} = A_{3,10} = A_{10,11}
	\end{array}\right\rbrace&=\Delta_{2},\\
	\left.\begin{array}{r}
	A_{2,3} = A_{3,1} = A_{5,4} = A_{4,6}\\
	A_{11,12} = A_{12,10} = A_{7,8} = A_{8,9}\\
	A_{1,5} = A_{5,12} = A_{6,2} = A_{2,8}\\
	A_{9,11} = A_{11,4} = A_{10,7} = A_{7,3}
	\end{array}\right\rbrace&=\eta_{0}\Delta_{2},
	\end{split}
	\end{equation}
	on the second NN bonds, and
	\begin{equation}\label{eq:ansatz23}
	\begin{split}
	A_{1,9}=A_{6,10} &= \Delta_{3},\\
	A_{2,11}=A_{4,3}=A_{5,7} = A_{8,12}&=\eta_{0}\Delta_{3},
	\end{split}
	\end{equation}
\end{subequations}
on the third NN bonds, where $\eta_{0}=\pm{}1$ and $\Delta_{1,2,3}$ are three complex numbers.
The gauge transformation given in Eq.~\eqref{eq:gauge} will change the signs of $\eta_{0}$ and $\Delta_{1}$ while leaving $\Delta_{2}$ and $\Delta_{3}$ unchanged, namely,  $$\eta_0 \longrightarrow -\eta_0,\,\, \Delta_{1}\longrightarrow -\Delta_{1}, \,\, \Delta_{2}\longrightarrow \Delta_{2},\,\, \Delta_{3}\longrightarrow \Delta_{3}.$$
So we can study the ansatzes with fixed $\eta_{0}$. Note that the time-reversal symmetry breaking is allowed in these zero-flux states.

In the next section, we shall study the $\pi$-flux state given in Eq.~\eqref{eq:ansatz1} and the zero-flux state given in Eq.~\eqref{eq:ansatz2} using mean-field theory.
It is noted that the $\pi$-flux state and the zero flux state with $\Delta_1=\Delta_3=0$ were used to study the $J_1$-$J_2$-$J_3$ model in the hyperkagome limit $J_1=J_3=0$ in Ref.~\onlinecite{NaIrO6}. The PSG analysis in this paper shows that they are only available $Z_2$ bosonic states in this limit. However, for generic situations with finite $J_1$ and $J_3$, there will be more allowed bosonic states in accordance with all the four PSGs given in Eq.~\eqref{eq:psg1}.

\section{Results of the Schwinger boson mean-field theory}\label{sec:results}
In this section, we study the ground states of the $J_1$-$J_2$-$J_3$ Heisenberg model based on the mean-field ansatzes given in Eqs.~\eqref{eq:ansatz1} and \eqref{eq:ansatz2}.

For the $\pi$-flux state given in Eq.~\eqref{eq:ansatz1}, there are only two independent parameters $\Delta_{2}$ and $\lambda$ in the mean-field ansatz, which are both real numbers. 

For the zero-flux state given in Eq.~\eqref{eq:ansatz2}, we can fix $\Delta_{2}$ to be real and write $\Delta_{1}$ and $\Delta_{3}$ in terms of their amplitudes and phases as follows:
\begin{equation}\label{eq:theta}
\Delta_{1}=|\Delta_{1}|e^{i\theta_{1}},\, \Delta_{3}=|\Delta_{3}|e^{i\theta_{3}},
\end{equation}
so that there are six independent real parameters $|\Delta_{1}|$, $\Delta_{2}$, $|\Delta_{3}|$, $\theta_1$, $\theta_3$ and $\lambda$ in the mean-field ansatz of the zero-flux state. In the calculations, we solve the self-consistent Eqs.~\eqref{eq:self} with fixed $\theta_{1}$ and $\theta_{3}$ to obtain the energy per site $E_g(\theta_1,\theta_3)$ for a local ground state. Then, we minimize $E_g$ with respect to $\theta_1$ and $\theta_3$ to reach the ground state with the energy minimum.

Note that the $\kappa$ defined in Eq.~\eqref{eq:kappa} serves as an additional control parameter in the calculations. For a given set of $J_1$, $J_2$, and $J_3$, there exists a critical value $\kappa_c$. When $\kappa<\kappa_c$, a gapped spin liquid state is favored, while when $\kappa>\kappa_c$, the spinon gap will close and the Schwinger bosons will condense at some gapless point in the $k$ space and give rise to a long-range magnetic order~\cite{MichaelMa89}. The details of the self-consistent calculation can be found in Appendix~\ref{ap:selfcon}, and the calculations of spin correlation function and spin static structure factor can be found in Appendix~\ref{ap:ssf}.

\subsection{$\kappa>\kappa_{c}$}
When $\kappa>\kappa_{c}$, the self-consistent equation will give rise to a gapless state where the Schwinger bosons will condensed at some gapless points $\{\bm{K}_1,\bm{K}_2,\cdots\}$, resulting in a magnetically ordered state. 
The magnetic orders are characterized by the peaks at wave vectors $\bm{Q}$ in the spin static structure factor $S(\bm{q})$. The allowed ordering wave vectors are given by
\begin{equation}\label{eq:Q}
\bm{Q}=\bm{K}_{i}-\bm{K}_{j},
\end{equation}
although the spin static structure factor $S(\bm{q})$ may vanish at some $\bm{Q}$ given in Eq.~\eqref{eq:Q} due to coherence factors (see Appendix~\ref{ap:ssf} for details). Usually we find that $\kappa_c<1$ for both the $\pi$-flux state and the zero-flux state, which can be also seen in some examples later. Thus the boson condensation will occur in these SU(2) spin-1/2 systems, where $S=1/2$ and $N=1$, and  thereby $\kappa=1$.

\begin{figure}[tbp]
	\includegraphics[width=8.4cm]{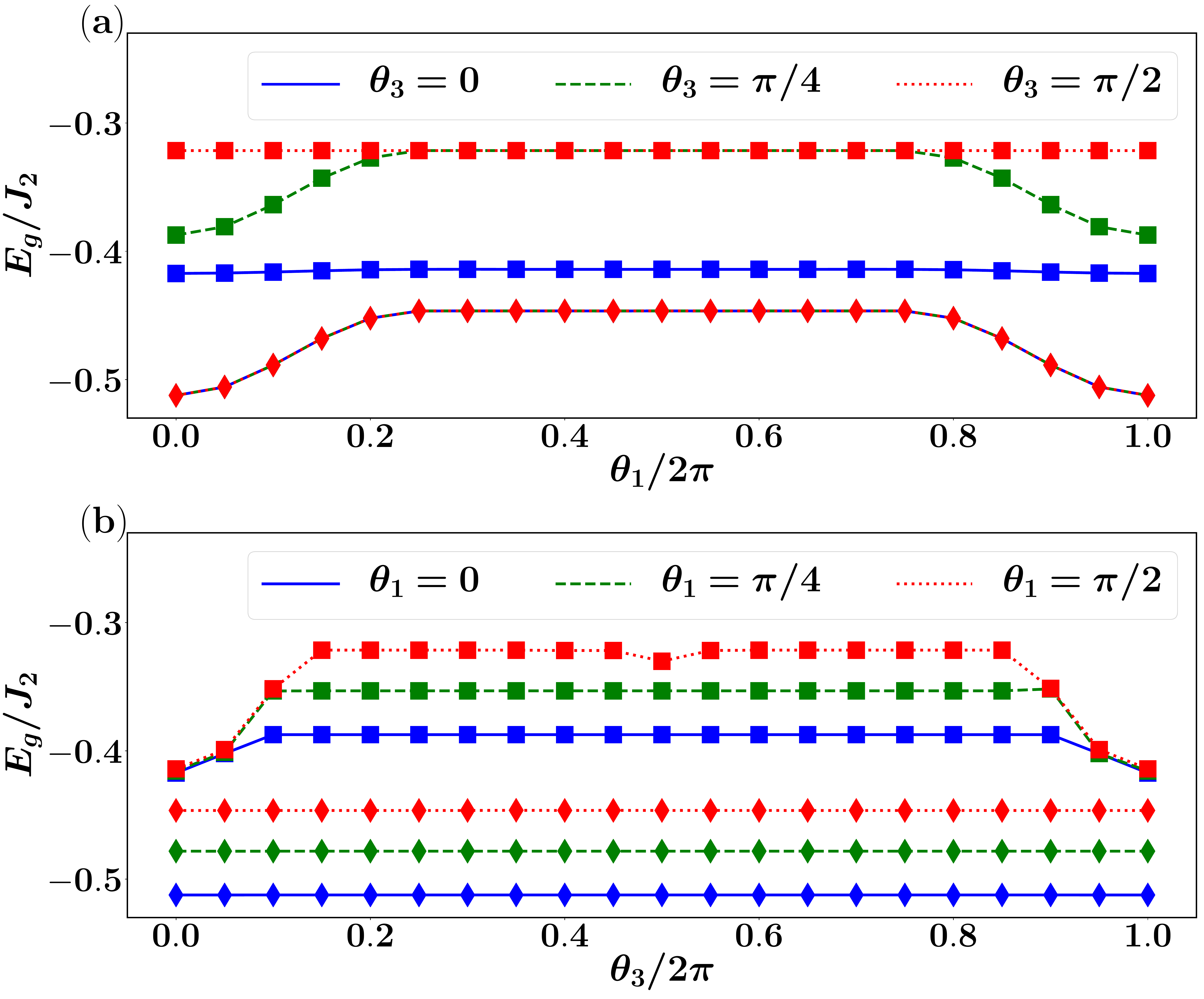}
	\caption{Zero-flux state: ground-state energy $E_{g}$ in the unit of $J_2$ is plotted versus (a) $\theta_1$ and (b) $\theta_3$. Here we set $J_{1}/J_{2}=0.3$, and the diamonds denote $J_{3}/J_{2}=0.3$ and the squares denote $J_{3}/J_{2}=0.8$. The energy minimum is achieved at $\theta_1=\theta_3=0$. When $J_{3}/J_{2}=0.3$, the amplitude $|\Delta_{3}|$ is tiny such that $E_{g}$ hardly changes with $\theta_{3}$. Therefore, the three curves (marked by diamonds) with $\theta_{3}=0,\pi/4,\pi/2$ almost collapse to a single one in panel (a), and become nearly flat in panel (b). More details can be found in Appendix~\ref{ap:d123}.} \label{fig:flux}
\end{figure}

For all the values of $J_1/J_2$ and $J_3/J_2$, we find that the zero-flux state always has lower energy than the $\pi$-flux state, and the energy minimum of a zero-flux state is always achieved at $\theta_1=\theta_3=0$, as demonstrated in Fig.~\ref{fig:flux}. Below we shall discuss the $\pi$-flux and the zero-flux states respectively.

\emph{$\pi$-flux state.---}
For the $\pi$-flux state, the mean-field order parameters vanish on the first and the third NN bonds, say, $\Delta_{1}=\Delta_{3}=0$. We find that the spinons will always condense at the following three time-reversal invariant wave vectors,
\begin{equation}\label{eq:pi-K}
\bm{K}_{1}=(0,\pi,\pi),\, \bm{K}_{2}=(\pi,0,\pi),\, \bm{K}_{3}=(\pi,\pi,0),
\end{equation}
where are related to each other by the $C_{3}$ rotations in space group $P4_132$. Thus, the static structure factor $S(\bm{q})$ will be peaked at $\bm{q}=\bm{Q}=\bm{K}_i-\bm{K}_j$. To be explicit, the wave vectors $\bm{Q}$ are given as follows:
\begin{equation}\label{eq:pi-Q}
\bm{Q}=(0,0,0),\ (\pi,\pi,0),\ (\pi,0,\pi),\ (0,\pi,\pi).
\end{equation}
Note that $S(\bm{Q}=0)$ for a $\pi$-flux state is finite in general when the time reversal symmetry is respected (see  Appendix~\ref{ap:ssf} for details).

\begin{figure}[tbp]
	\includegraphics[width=8.4cm]{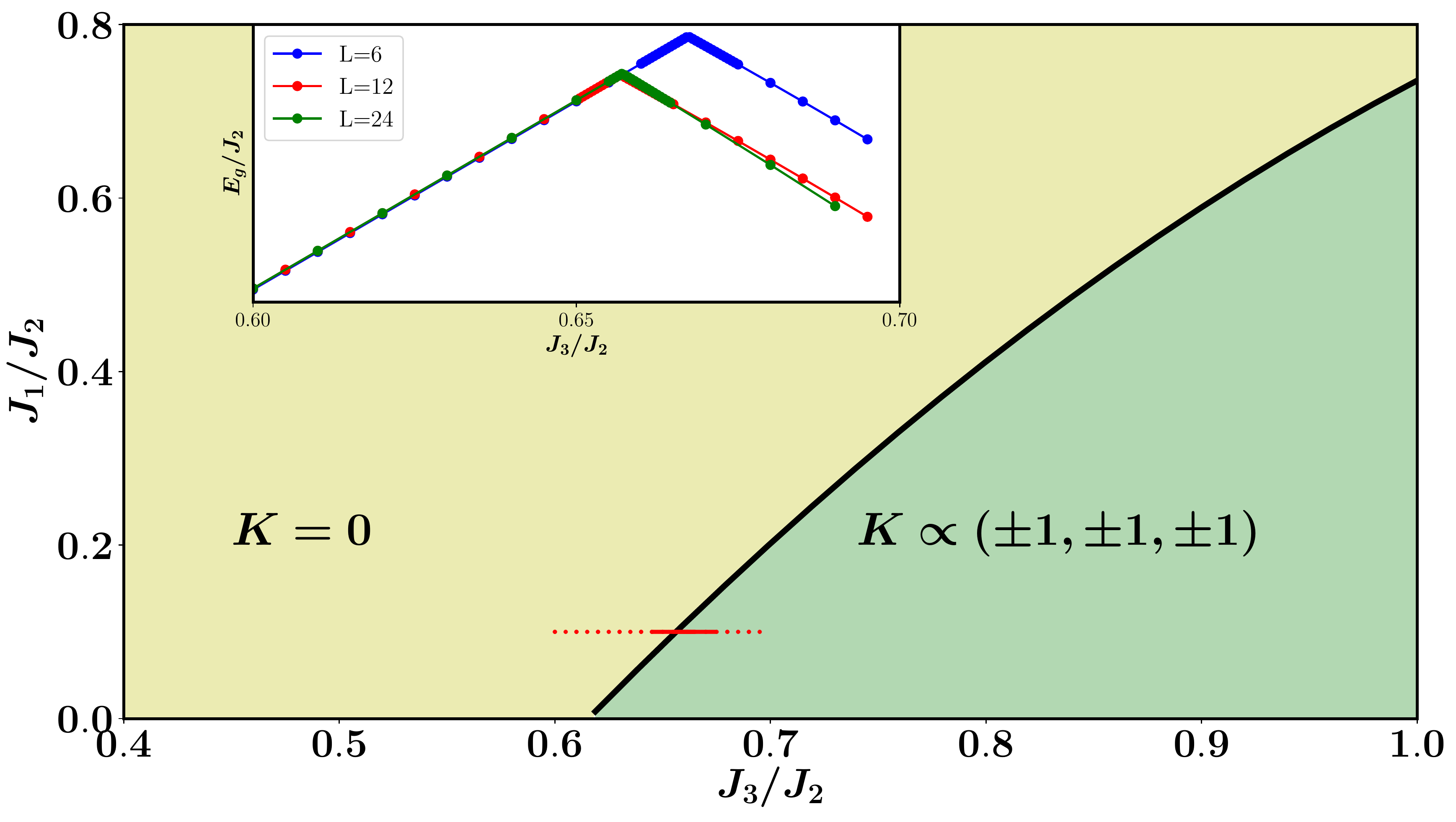}
	\caption{The ground-state phase diagram for $\kappa>\kappa_c$. The ground states are all zero-flux states with Schwinger boson condensation at wave vector $\bm{K}$. There are two phases found: (1) a uniform phase with $\bm{K}=0$ and (2) an incommensurate phase with $\bm{K}=K_0(\pm 1,\pm1, \pm1)$. Here $J_{1}=J_{tri}$, $J_{2}=J_{hyper}$, and $J_{3}=J_{chain}$ are the AFM Heisenberg exchanges on triangle bonds, hyperkagome bonds, and uniform chain bonds respectively. Insert: The ground-state energy $E_g$ vs $J_{3}/J_{2}$ along the line $J_{1}/J_{2}=0.1$. The red dots crossing the phase boundary denote the parameter regions for the insert plot. }\label{fig:phase}
\end{figure}

\emph{Zero-flux state.---}
For the zero-flux state, we always have $\theta_1=\theta_3=0$ for the ground state, as demonstrated in Fig.~\ref{fig:flux}. 
It is also found that the zero-flux state is energetically favored in comparison with the $\pi$-flux state. 
We change the ratios $J_1/J_2$ and $J_3/J_2$ to explore the phase diagram. As plotted in Fig.~\ref{fig:phase}, there exist two distinct phases belonging to the same class of ansatz in accordance with the PSG. 
One is a uniform state with boson condensation at $\bm{K}=0$. The other is an incommensurate phase (see Appendix~\ref{ap:incomm} for details) with boson condensation at
\begin{equation}\label{eq:zero-K}
\bm{K}=K_0(\pm{}1, \pm1, \pm1),
\end{equation}
where $K_{0}/2\pi$ takes the values of $0.42\sim{}0.5$. For a uniform state, the static structure factor $S(\bm{q})$ has only one peak at $\bm{Q}=0$; while for an incommensurate state, $S(\bm{q})$ has peaks at multiple $\bm{Q}$'s as follows:
\begin{eqnarray}\label{eq:zero-Q}
\bm{Q}&=& 2K_0(\pm 1,\pm 1, \pm 1),\, 2K_0(\pm 1,0, 0), \nonumber\\
&& 2K_0(0,\pm 1, 0),\, 2K_0(0,0,\pm 1).
\end{eqnarray}
The peaks at other $\bm{Q}=\bm{K}_{i}-\bm{K}_{j}$ will vanish because the condensation wave vectors $\bm{K}$ in Eq.~\eqref{eq:zero-K} are not time reversal invariant and not degenerate, such that the coherence factors are destructive and give rise to vanishing $S(\bm{Q})$ (see Appendix~\ref{ap:ssf} for details).

It is worth mentioning that the wave vector $K_0$ is incommensurable in general, which is similar to that in the classical model, and the details for the incommensurable $K_0$ can be found in Appendix~\ref{ap:incomm}. It is also noted that the incommensurate states break the lattice translational symmetry because of the boson condensation, even though the corresponding mean-field ansatz is translationally invariant. There exists a phase transition separating the uniform states from the incommensurate states.
The transition between the uniform phase and the incommensurate phase is of first order, which can be viewed from the insert plot in Fig.~\ref{fig:phase}. The kink at the phase boundary gives rise to a discontinuous jump in the first-order derivative $\partial E_g/\partial J_{3}$, and suggests a first-order phase transition. Meanwhile, the wavevector  $K_0/2\pi$ jumps from $0$ to $\sim 0.42$, crossing the phase boundary.

\subsection{$\kappa<\kappa_{c}$}
When $\kappa<\kappa_{c}$, the ground state is a spin liquid state with bosonic spinons, where the spinon excitation spectrum has a finite energy gap. Since a single spinon carries a $U(1)$ gauge field and is not gauge invariant, we would like to consider the two-spinon excitations, which are physically measurable in a spinon-deconfined state. The energy of such a two-spinon excitation is given by
\begin{equation}
\begin{split}
\omega_{2s,\mu\nu}(\bm{k}) = \omega_{\mu}(\bm{k}_{1}) + \omega_{\nu}(\bm{k}_{2})\, \mbox{ with }\, \bm{k}=\bm{k}_{1}+\bm{k}_{2},
\end{split}
\end{equation}
where $\omega_{\mu}(\bm{k})$ is the $\mu$th band spinon dispersion. For a given $\bm{k}$, there exist infinite pairs of $\bm{k}_1$ and $\bm{k}_2$ that satisfy the momentum conservation relation $\bm{k}=\bm{k}_1+\bm{k}_2$, so that the spectrum of the two-spinon excitations is not characterized by sharp peaks in the dynamic spin structure factor $S(\bm{q},\omega)$, which defines the energy dispersion as the magnons in a magnetically ordered state. 
Instead, the two-spinon excitations form a continuous spectrum in $S(\bm{q},\omega)$, whose upper and lower edges can be measured by inelastic neutron scattering experiment. The lower edge of the two-spinon spectrum is given by
\begin{equation}
\omega_{2s}(\bm{k}) = \mbox{min}\{\omega_{1}(\bm{q}) + \omega_{1}(\bm{k}-\bm{q})\}, \label{eq:2spinon}
\end{equation}
where $\omega_{1}(\bm{k})$ is the lowest band single-spinon dispersion. 

We shall use the patterns of $\omega_{2s}(\bm{k})$ to characterize the $\pi$-flux phase and the zero-flux phase. In the calculations, we choose the parameters as $J_{1}/J_{2}=0.54$ and $J_{3}/J_{2}=0.77$, which were estimated by the density functional theory for relative exchange couplings between the Cu atoms in PbCuTe$_2$O$_6$~\cite{PbCuTe1}. For this set of $J_{1}/J_{2}$ and $J_{3}/J_{2}$, we find $\kappa_{c}=0.64$ for the zero-flux state and $\kappa_{c}=0.8$ for the $\pi$-flux state. For comparison, we choose $\kappa=0.4<\kappa_c$ to study gapful zero-flux and $\pi$-flux states.
Note that our calculations are done in the large-$N$ limit, where the Schwinger boson mean-field theory will be exact and the $O(1/N)$ and higher order corrections can be neglected. However, the physical realization is given by $N=1$, such that these corrections are considerable and $\kappa_{c}(N=1)$ deviates from $\kappa_{c}(N=\infty)$. Regarding the material PbCuTe$_2$O$_6$~\cite{PbCuTe1,PbCuTe2}, it could be either a quantum spin liquid state or a weakly ordering state with strong magnetic fluctuations.

\begin{figure}[tbp]
	\includegraphics[width=8.4cm]{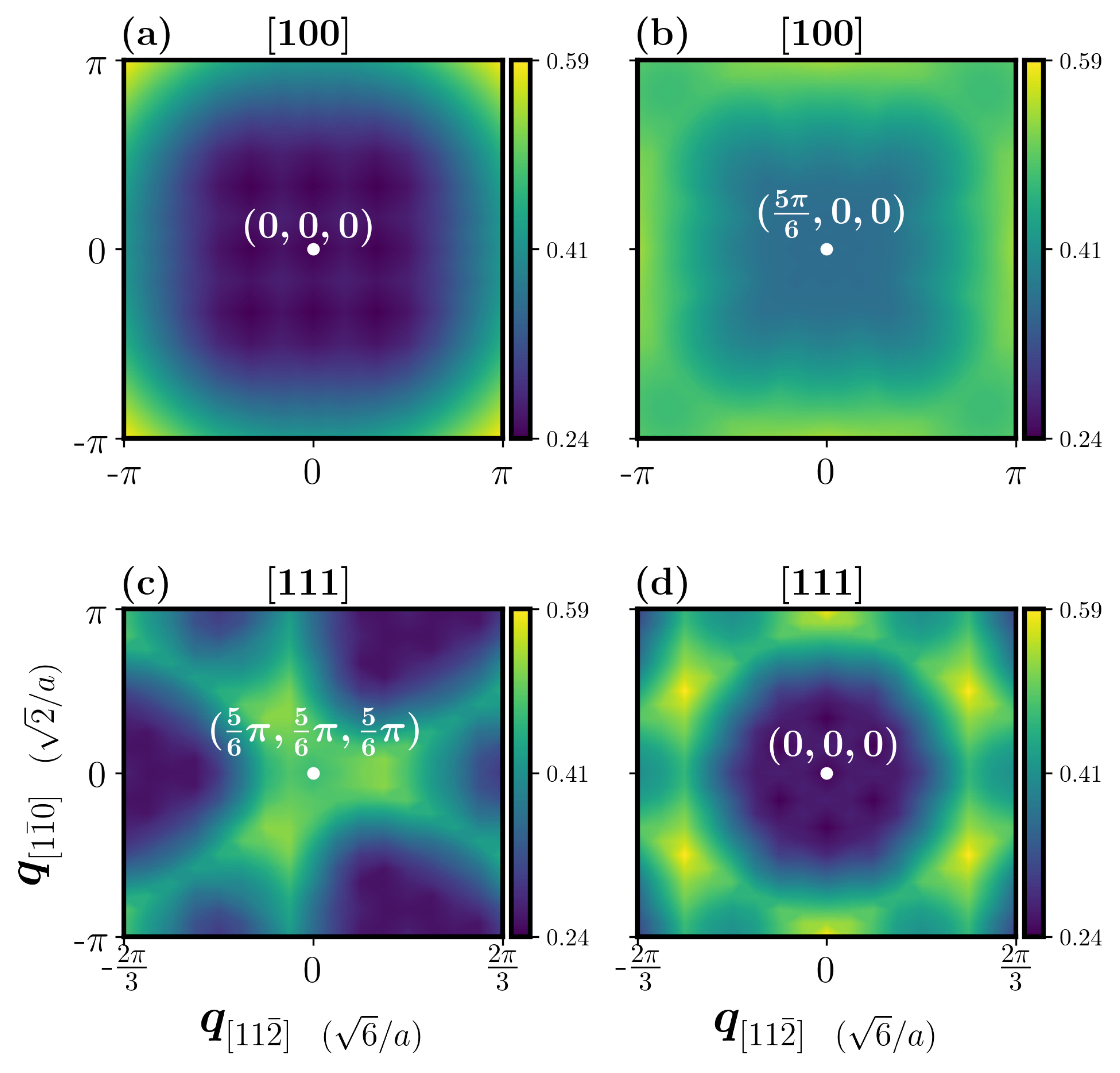}
	\caption{A zero-flux state: the lower edge of the two-spinon spectrum $\omega_{2s}(\bm{k})$ on two $[100]$ planes [(a) and (b)] and two $[111]$ planes [(c) and (d)]. The parameters are chosen as $\kappa=0.4$, $J_2=1$, $J_{1}=0.54$ and $J_{3}=0.77$. Two $[100]$ planes passing through (a) $(0,0,0)$ and (b) $(5\pi/6,0,0)$, and two $[111]$ planes passing through (c) $(5\pi/6,5\pi/6,5\pi/6)$ and (d) $(0,0,0)$ respectively.
}\label{fig:2szero}
\end{figure}

{\em Zero-flux state.} For a zero-flux state, there are eight minima in the lowest band spinon dispersion, $\omega_{1}(\bm{k})$, which locate at
\begin{subequations}
\begin{equation}
\bm{k}=k_{0}(\pm{}1,\pm{}1,\pm{}1),
\end{equation}
with $k_{0}\approx\frac{5\pi}{6}$.
Consequently, the two-spinon spectrum, $\omega_{2s,\mu\nu}(\bm{q})$, will reach the energy minima at
\begin{eqnarray}\label{eq:minima-zero}
\bm{q} & = & (0,0,0),\, q_{0}(\pm{1},\pm{1},\pm{1}), \nonumber\\
& & q_{0}(\pm{1},0,\pm{1}),\, q_{0}(0,\pm{1},\pm{1}),\, q_{0}(\pm{1},\pm{1},0), \nonumber\\
& & q_{0}(\pm{1},0,0),\, q_{0}(0,\pm{1},0),\, q_{0}(0,0,\pm{1}),
\end{eqnarray}
where $q_{0}\approx \frac{5\pi}{3}\equiv -\frac{\pi}{3}(\mbox{mod} 2\pi)$.
\end{subequations}
The lower edge of the two-spinon spectrum, $\omega_{2s}(\bm{k})$, for the zero-flux state is shown in Fig.~\ref{fig:2szero}, which exhibits fourfold rotational symmetry along the [100] direction and threefold rotational symmetry along the [111] direction. 

\begin{figure}[tbp]
	\includegraphics[width=8.4cm]{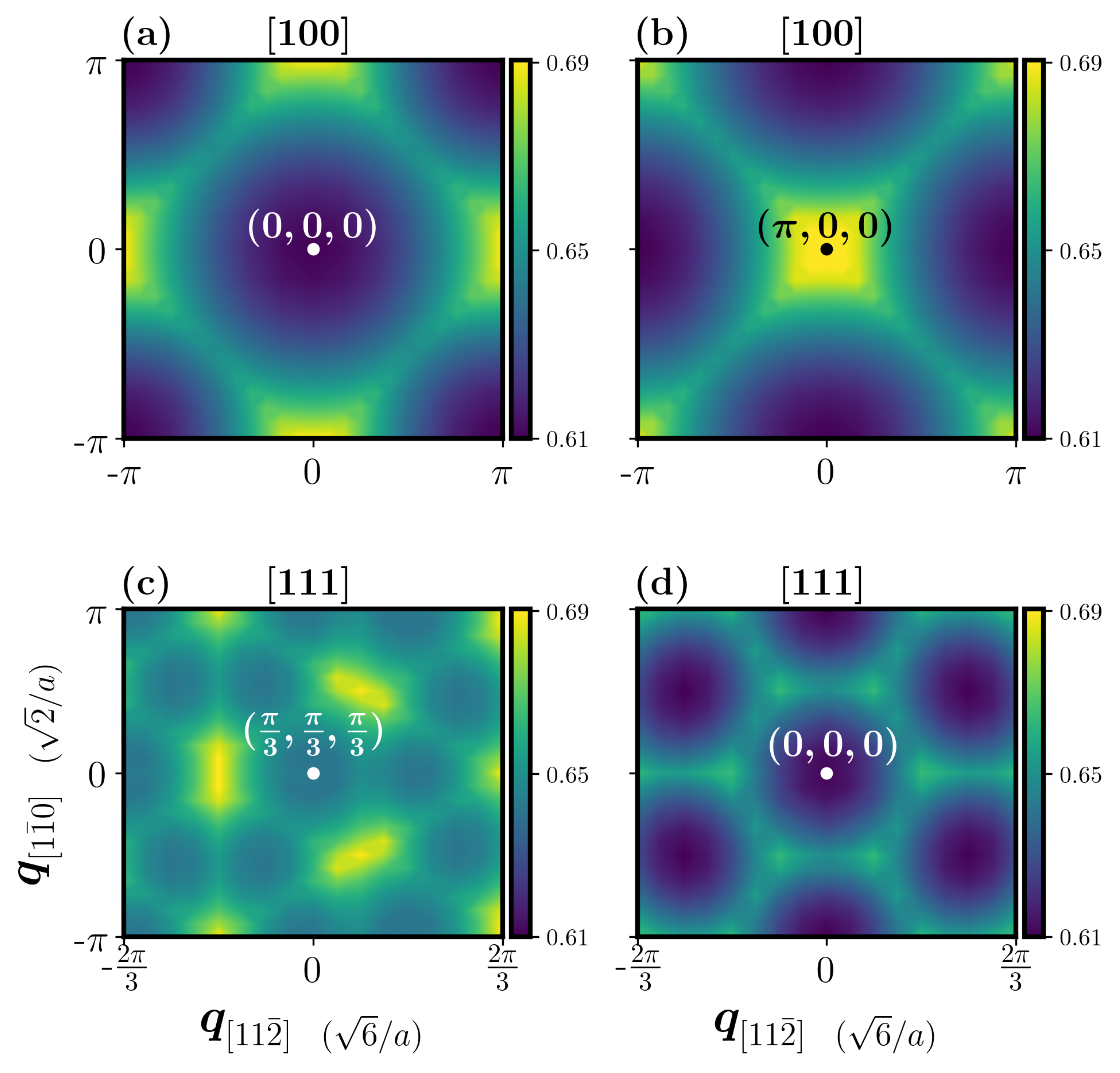}
	\caption{A $\pi$-flux state:  the lower edge of the two-spinon spectrum $\omega_{2s}(\bm{k})$ on two $[100]$ planes [(a) and (b)] and two $[111]$ planes [(c) and (d)]. The parameters are chosen as $\kappa=0.4$, $J_2=1$, $J_{1}=0.54$ and $J_{3}=0.77$. Two $[100]$ planes passing through (a) $(0,0,0)$ and (b) $(\pi,0,0)$, and two $[111]$ planes passing through (c) $(\pi/3,\pi/3,\pi/3)$ and (d) $(0,0,0)$ respectively.
	}\label{fig:2spi}
\end{figure}

{\em $\pi$-flux state.} It is found that ground state energy of the $\pi$-flux state is higher than the one of the zero-flux state for the same set of $\{J_1,J_2,J_3,\kappa\}$. For the $\pi$-flux state, there are three minima in the lowest band spinon dispersion $\omega_{1}(\bm{k}_{0})$, which locate at the three time-reversal invariant points as follows:
\begin{subequations}
\begin{equation}
\bm{k}=(\pi,\pi,0),\, (\pi,0,\pi),\, (0,\pi,\pi).
\end{equation}
Therefore, the energy minima of the two-spinon spectrum will occur at
\begin{equation}\label{eq:minima-pi}
\begin{split}
\bm{q}=(0,0,0),\, (\pi,\pi,0),\, (\pi,0,\pi),\, (0,\pi,\pi).
\end{split}
\end{equation}
\end{subequations}
The lower edge of the two-spinon spectrum for the $\pi$-flux state is shown in Fig.~\ref{fig:2spi}, which exhibits fourfold rotational symmetry along the [100] direction and threefold rotational symmetry along the [111] direction, as the same as the zero-flux state. These fourfold and threefold symmetries are consistent with $S_4$ and $C_3$ symmetry operations in lattice $P4_132$ space group. 

\section{Discussions and conclusions}\label{sec:summary}

To summarize, in this paper we have studied hyperkagome lattice $J_1$-$J_2$-$J_3$ AFM Heisenberg model in both classical and quantum limits, where $J_1=J_{tri}$ is the first NN AFM coupling on isolated triangle bonds, $J_2=J_{hyper}$ is the second NN AFM coupling on hyperkagome bonds, and $J_3=J_{chain}$ is the third NN AFM coupling along the uniform chains.

In the classical limit, we have analyzed the classical $J_{1}$-$J_{2}$-$J_{3}$ Heisenberg model with the help of two triangle rules, the hyperkagome triangle rule (associated with $J_2$) and the isolated triangle rule (associated with $J_1$), and explored the whole phase diagram by Monte Carlo simulations. There is a total of five phases found (see Fig.~\ref{fig:cphase}): (1) a disordered state with huge residual entropy down to zero temperature at $J_1=J_3=0$, whose ground states are subject to the hyperkagome triangle rule; (2) an ordered phase locates along the axis $J_3=0$ and $J_1>0$, where two types of coplanar ordered states $\bm{Q}=0$ and $\bm{Q}=\frac{2\pi}{3}(\pm 1,\pm 1,\pm 1)$ are degenerate (these coplanar states satisfy both the hyperkagome triangle rule and the isolated triangle rule); (3) a noncoplanar phase with $\bm{Q}=0$ exists when $J_1\gg J_3$; (4) a noncoplanar phase with $\bm{Q}=\frac{2\pi}{3}(\pm 1,\pm 1,\pm 1)$ exists when $J_1\ll J_3$; and (5) a noncoplanar incommensurate state with wave vector $\bm{Q}\propto{}(\pm{}1,\pm{}1,\pm{}1)$ appears when $J_1$ and $J_3$ are comparable. All the noncoplanar states violate both the hyperkagome triangle rule and the isolated triangle rule. It is worth mentioning that the results for $J_1=0$ or $J_3=0$ are exact, and the corresponding spin configurations on the ground states are given in Eqs.~\eqref{eq:config1}, \eqref{eq:config2}, and \eqref{eq:config3}.

In the quantum regime, we have applied the Schwinger boson representation (with $2N$ species of bosons) and the large-$N$ expansion to formulate the $J_{1}$-$J_{2}$-$J_{3}$ Heisenberg model, which allows us to study both quantum spin liquid states and long-range magnetically ordered states on an equal footing. 
In this formulation, the physical properties are controlled by the ratio $\kappa=n_{b}/N$, where $n_{b}$ is the number of boson at each site (spin). There exists a critical value $\kappa_c$: When $\kappa>\kappa_c$, the condensation of Schwinger bosons will happen and give rise to a long-range magnetic order, while when $\kappa<\kappa_c$, the bosonic spinons have a finite excitation gap and lead to a gapful quantum spin liquid state. 
Note that the magnetically ordered states obtained by Schwinger boson condensation can also be described by the spin wave theory when $N=1$.

We have classified these (gapless or gapped) bosonic sates using PSG in accordance with the lattice $P4_{1}32$ symmetry. It is found that the non-symmorphic space group $P4_{1}32$ imposes strong constraints on SU(2) symmetry fractionalization, and there exist only four A-type (associated with translationally invariant mean-field ansatzes) $Z_{2}$ bosonic algebraic PSGs. Considering the concrete $J_{1}$-$J_{2}$-$J_{3}$ Heisenberg model, where only the first three NN bonds are nonzero, there are only two types of bosonic states are allowed. These two types of bosonic states are distinct from each other by the gauge-invariant flux, $\Phi_{10}$, on the elementary 10-site loops on the hyperkagome network. One has $\Phi_{10}=0$ and is called the zero flux state; the other has $\Phi_{10}=\pi$ and is called the $\pi$-flux state. 

For the zero-flux state, the mean-field order parameters on all the three first NN bonds are nonzero in general, while for the $\pi$-flux state, the mean-field order parameters vanish on the first and the third NN bonds, and only the second NN bonds (hyperkagome bonds) are nonzero. It means that the zero-flux state has more order parameters than the $\pi$-flux state in the mean-field ansatz. 
The self-consistent mean-field theory calculation for the $J_{1}$-$J_{2}$-$J_{3}$ Heisenberg model on hyperkagome lattices finds that the zero-flux state is energetically favored rather than the $\pi$-flux state.

Both the zero-flux states and $\pi$-flux states are able to give rise to gapful quantum spin liquid state when $\kappa<\kappa_c$ and long-range magnetic order when $\kappa>\kappa_c$. In both situations, $\kappa>\kappa_c$ and $\kappa<\kappa_c$, the zero-flux states can be distinguished from the $\pi$-flux states experimentally. (i) When $\kappa>\kappa_c$, the gapless $\pi$-flux state will give rise to four peaks in the spin static structure factor $\bm{S}(\bm{Q})$ at the time-reversal invariant wave vectors $\bm{Q}=(0,0,0)$, $(0,\pi,\pi)$, $(\pi,0,\pi)$, and $(\pi,\pi,0)$ respectively, while the zero-flux state either is uniform and has only one $\bm{S}(\bm{Q})$ peak at $\bm{Q}=(0,0,0)$, or is incommensurate and has a set of peaks at $\bm{Q}=Q_0(\pm 1,\pm 1,\pm 1)$, $Q_0(\pm 1,0,0)$, $Q_0(0,\pm 1,0)$, and $Q_0(0,0,\pm 1)$. The zero-flux state phase diagram for $\kappa>\kappa_{c}$ is plotted in Fig.~\ref{fig:phase}. (ii) When $\kappa<\kappa_c$, the ground states are quantum spin liquid states, and the elementary spin excitations are bosonic spinons that have a finite excitation gap. The spin spectral function $S(\bm{q},\omega)$ is no longer characterized by the sharp spin-wave dispersions as in the magnetically ordered states. Instead, $S(\bm{q},\omega)$ will display a broad spinon continuum, which measures the two-spinon spectrum indeed. The patterns of the lower edges of the two-spinon spectrum can be used to distinguish the $\pi$-flux state from the zero-flux state. For the zero-flux state, the two-spinon spectrum has energy minima at $\bm{q}=(0,0,0)$, $q_0(\pm 1, \pm 1,\pm1)$, $q_0(0,\pm 1,\pm 1)$, $q_0(\pm 1,0,\pm 1)$, $q_0(\pm 1,\pm 1,0)$, $q_0(\pm 1, 0,0)$, $q_0(0,\pm 1,0)$, and $q_0(0,0,\pm 1)$, where $q_0$ is incommensurable generally, while for the $\pi$-flux state, the two-spinon spectrum has energy minima at $\bm{q}=(0,0,0)$, $(0,\pi,\pi)$, $(\pi,0,\pi)$, and $(\pi,\pi,0)$.

We hope the studies in this paper will shed light on future experiments on PbCuTe$_2$O$_6$ and other possible hyperkagome antiferromagnets.

\section{acknowledgement}

We would like to thank Jun Zhao for helpful discussions and for sharing the experimental data on PbCuTe$_2$O$_6$ before publication. The communications and discussions with Yuan-Ming Lu and Biao Huang are acknowledged. 
This work is supported in part by National Key Research and Development Program of China (No.2016YFA0300202),
National Natural Science Foundation of China (No. 11774306), and the Strategic Priority Research Program of Chinese Academy of Sciences (No. XDB28000000).

\appendix
\begin{appendices}

\section{Self-consistent equations in Schwinger boson mean-field theory}\label{ap:selfcon}
In this section, we formulate the self-consistent equations in Schwinger boson mean-field theory for $Sp(N=1)$, and the formulation for generic $N$ can be established in a similar way. Here we use the $J_{1}$-$J_{2}$-$J_{3}$ Heisenberg model on a hyperkagome lattice as an example for study.

By performing Fourier transformation, we introduce the $k$-space Schwinger bosons as
\begin{equation}
\begin{split}
&\bm{b}_{\bm{k}\alpha} = \left(b_{\bm{k}1\alpha},\dots,b_{\bm{k}12\alpha}\right),\quad\alpha=\uparrow,\downarrow,\\
\end{split}
\end{equation}
where $\bm{k}$ is a three-dimensional vector in the first Brillouin zone. For a given real-space mean-field ansatz $A_{i\mu{}j\nu}$, the Hamiltonian in Eq.~\eqref{eq:HMF} can be rewritten as
\begin{equation}
\begin{split}
&\mathcal{H}_{MF} = \sum_{a=1}^{3}\sum_{\langle{}i\mu,j\nu\rangle_{a}}2J_{a} |A_{i\mu{}j\nu}|^{2}\\
&\qquad\qquad\qquad + \text{C}- 12\lambda{}N_{u}(1+\kappa) + \mathcal{H}_{b},\\
&\mathcal{H}_{b} =\sum_{\bm{k}}\Psi(\bm{k})^{\dagger}D(\bm{k})\Psi(\bm{k})
\end{split}
\end{equation}
where $C$=$\sum_{a=1}^{3}\sum_{\langle{}i\mu,j\nu\rangle_{a}}J_{a}/8$ is a constant, $N_{u}$ is the number of unit cell, and the vector spinon field $\Psi(\bm{k})$ and the $24\times{}24$ matrix $D(\bm{k})$ are
\begin{equation}
\begin{split}
\Psi(\bm{k})& = \left(\begin{array}{c}
\bm{b}_{\bm{k}\uparrow} \\ \bm{b}_{-\bm{k}\downarrow}^{\dagger}
\end{array}\right),\\
D(\bm{k})&=\left(\begin{array}{cc}
\Lambda          & -A_{\bm{k}}  \\
-A_{\bm{k}}^{\dagger} & \Lambda
\end{array}\right).
\end{split}
\end{equation}
Here $\Lambda$ and $A_{\bm{\bm{k}}}$ are $12\times{}12$ matrices and read 
\begin{equation}
\begin{split}
&\Lambda_{\mu\nu} = \lambda\delta_{\mu\nu},\\
&(A_{\bm{k}})_{\mu\nu} = \frac{1}{N_{u}}\sum_{a=1}^{3}\sum_{\langle{}i\mu,j\nu{}\rangle_{a}}J_{a}A_{i\mu{}j\nu}e^{i\bm{k}\cdot{}(\bm{r}_j-\bm{r}_i)},
\end{split}
\end{equation}
where $\bm{r}_{i}$ denotes the position vector of unit cell $i$, and $\langle{}i\mu,j\nu{}\rangle_{a}$ denotes the $a$th NN bonds. Recall that the explicit forms of $A_{i\mu{}j\nu}$ are defined in Eqs.~\eqref{eq:ansatz1} and \eqref{eq:ansatz2}. 
Following the notation in the main text, we define $|A_{i\mu{}j\nu}|\equiv|\Delta_{a}|$($a=1,2,3$) on the $a$th NN bond.

By performing singular value decomposition (SVD) 
\begin{equation}\label{eq:svd}
A_{\bm{k}} = U_{\bm{k}}E_{\bm{k}}V_{\bm{k}}^{\dagger},
\end{equation} 
we obtain
\begin{equation}
\begin{split}
&\mathcal{H}_{b} =\sum_{\bm{k}\mu}\left(
\tilde{b}_{\bm{k}\mu\uparrow}^{\dagger}\  \tilde{b}_{-\bm{k}\mu\downarrow}
\right)h_{\mu}(\bm{k})\left(
\begin{array}{c}
\tilde{b}_{\bm{k}\mu\uparrow} \\ \tilde{b}_{-\bm{k}\mu\downarrow}^{\dagger}
\end{array}\right),\\
&h_{\mu}(\bm{k}) = \left(\begin{array}{cc}
\lambda & -(E_{\bm{k}})_{\mu}\\
-(E_{\bm{k}})_{\mu} & \lambda
\end{array}\right),
\end{split}
\end{equation}
where
\begin{equation}
\begin{split}
&\tilde{\bm{b}}_{\bm{k}\alpha} = \left(\tilde{b}_{\bm{k}1\alpha},\dots,\tilde{b}_{\bm{k}12\alpha}\right),\quad\alpha=\uparrow,\downarrow,\\
&\bm{b}_{\bm{k}\uparrow} = U_{\bm{k}}\tilde{\bm{b}}_{\bm{k}\uparrow},\qquad{}\bm{b}_{-\bm{k}\downarrow}^{\dagger} = V_{\bm{k}}\tilde{\bm{b}}_{-k\downarrow}^{\dagger}.
\end{split}
\end{equation}
Because $A_{\bm{k}}^{\dagger}=-A_{\bm{k}}$ (i.e., $iA_{\bm{k}}$ is Hermitian), both $U_{\bm{k}}$ and $V_{\bm{k}}$ are unitary matrices. The SVD allows us to block diagonalize the $24\times{}24$ matrix $D(\bm{k})$ into twelve $2\times{}2$ matrices $h_{\mu}(\bm{k})$($\mu=1,...,12$), and then each $2\times{}2$ matrix can be diagonalized independently. With the help of Bogoliubov transformation as 
\begin{equation}
\begin{split}\label{eq:gamma2b}
&b_{\bm{k}\mu\uparrow}=\sum_{\nu}(U_{\bm{k}})_{\mu\nu}(u_{\bm{k}\nu}\gamma_{\bm{k}\nu\uparrow}-v_{\bm{k}\nu}\gamma_{-\bm{k}\nu\downarrow}^{\dagger}),\\
&b_{-\bm{k}\mu\downarrow}^{\dagger}=\sum_{\nu}(V_{\bm{k}})_{\mu\nu}(-v_{\bm{k}\nu}\gamma_{\bm{k}\nu\uparrow}+u_{\bm{k}\nu}\gamma_{-\bm{k}\nu\downarrow}^{\dagger}),\\
\end{split}
\end{equation}
we can obtain 
\begin{equation}
\mathcal{H}_{b} = \sum_{\bm{k}\mu}\omega_{\mu}(\bm{k})(\gamma^{\dagger}_{\bm{k}\mu\uparrow}\gamma_{\bm{k}\mu\uparrow}+\gamma^{\dagger}_{\bm{k}\mu\downarrow}\gamma_{\bm{k}\mu\downarrow} + 1),
\end{equation}
where $\omega_{\mu}(\bm{k})$ is the $\mu$th band single spinon dispersion as
\begin{equation}
\begin{split}
&\omega_{\mu}(\bm{k}) = \sqrt{\lambda^{2}-|(E_{\bm{k}})_{\mu}|^{2}},\\
\end{split}
\end{equation}
and
\begin{equation}
\begin{split}\label{eq:uv}
&u_{\bm{k}\mu}=\sqrt{\frac{\lambda}{2\omega_{\mu}(\bm{k})}+\frac{1}{2}},\\
&v_{\bm{k}\mu}=-\sqrt{\frac{\lambda}{2\omega_{\mu}(\bm{k})}-\frac{1}{2}},\\
&|u_{\bm{k}\mu}|^{2}-|v_{\bm{k}\mu}|^{2}=1.\\
\end{split}
\end{equation}
The mean-field ground state energy reads
\begin{equation}
E_{g} =N_{u}\sum_{a=1}^{3}M_{a}J_{a}|\Delta_{a}|^{2}
- 12\lambda{}N_{u}(1+\kappa) + \sum_{\bm{k}\mu}\omega_{\mu}(\bm{k}),
\end{equation}
where $M_{1}=M_{2}/2=M_{3}=48$, and the constant C is omitted.
Below we will discuss how to determine $\Delta_{a}$ ($a=1,2,3$) self-consistently for both gapped states and gapless states.

{\em Gapped states.---} 
For a gapped state, $\omega_{\mu}(\bm{k})>0$, such that the matrix $h_{\mu}(\bm{k})$ is positive definite for any given $(\bm{k},\mu)$. Then the matrix $h_{\mu}(\bm{k})$ can be diagonalized by finite values of $u_{\bm{k}\mu}$ and $v_{\bm{k}\mu}$, and the corresponding self-consistent equations read
\begin{equation}
\begin{split}
& 12N_{u}(1+\kappa) = \sum_{\bm{k}\mu}\left(|u_{\bm{k}\mu}|^{2} + |v_{\bm{k}\mu}|^{2}\right),\\
&N_{u}M_{a}J_{a}|\Delta_{a}|^{2} = \sum_{\bm{k}\mu}2\tilde{A}^{a}_{\bm{k}\mu}v_{\bm{k}\mu}^{*}u_{\bm{k}\mu},\  a = 1,2,3
\end{split}
\end{equation}
where
\begin{equation}
\tilde{A}^{a}_{\bm{k}\mu}=\sum_{\langle{}\mu^{\prime}\nu^{\prime}\rangle_{a}}(U^{\dagger}_{\bm{k}})_{\mu\mu^{\prime}}(A_{\bm{k}})_{\mu^{\prime}\nu^{\prime}}(V_{\bm{k}})_{\nu^{\prime}\mu},\  a = 1,2,3,
\end{equation}
and $\langle{}\mu\nu\rangle_{a}$ denotes the $a$th NN bond which are form by sublattice $\mu$ and $\nu$.

{\em Gapless states.}
When there exist gapless points $\bm{k}$'s, in which $\omega_{\mu}(\bm{k})=0$, the boson condensation arises and the zero modes (the condensate part)  must be handled carefully, because the nonpositive matrix $h_{\mu}(\bm{k})$ corresponding to the zero modes can not be diagonalized by finite $u_{\bm{k}\mu}$ and $v_{\bm{k}\nu}$. We use $\alpha$ to denote the percentage of the condensate fraction, and the explicit form of the constraint Eq.~\eqref{eq:cstrnt} reads
\begin{equation}
\begin{split}
& 12N_{u}(1+\kappa) = \sum_{\omega_{\bm{k}\mu}\neq{}0}\left(|u_{\bm{k}\mu}|^{2} + |v_{\bm{k}\mu}|^{2}\right) + 12\alpha{}N_{\mu},\\
\end{split}
\end{equation}
and because of the gapless points, $\lambda$ can be straightforwardly obtained as 
\begin{equation}
\lambda = \max\limits_{\bm{k}\mu}\{(E_{\bm{k}})_{\mu}\}.
\end{equation}
Eventually, for gapless states, there remain only three self-consistent equations which read
\begin{equation}
\begin{split}
&N_{u}M_{a}J_{a}|\Delta_{a}|^{2} = \sum_{\omega_{\bm{k}\mu}\neq{}0}2\tilde{A}^{a}_{\bm{k}\mu}v_{\bm{k}\mu}^{*}u_{\bm{k}\mu}\\&\qquad+\sum_{\omega_{\bm{k}\mu=0}}\alpha{}N_{u}|\tilde{A}^{a}_{\bm{k}\mu}|,\quad a=1,2,3.
\end{split}
\end{equation}

\section{PSGs with space group P4$_{1}$32}\label{ap:PSG}
This appendix is devoted to derive the allowed bosonic algebraic PSGs for symmetric spin states on the hyperkagome lattice $P4_132$. At first, the constraints between symmetry group operators should be clarified.
The algebraic relations among $P4_132$ generators, $S_4$, $C_2$ and $C_3$, and the lattice translations $T_1$, $T_2$ and $T_3$ are 
{\small
\begin{align}
& C_{2}^{2}=C^{3}_{3} = T^{-1}_{1}S^{4}_{4} = I,                                                            \\
& T^{-1}_{1}C_{2}T^{-1}_{1}C_{2} = T^{-1}_{2}C_{2}T^{-1}_{3}C_{2}	=T^{-1}_{3}C_{2}T^{-1}_{2}C_{2} = I,      \\
& T^{-1}_{1}C_{3}^{-1}T_{2}C_{3} = T^{-1}_{2}C_{3}^{-1}T_{3}C_{3}	=T^{-1}_{3}C_{3}^{-1}T_{1}C_{3} = I,      \\
& T^{-1}_{1}S_{4}^{-1}T_{1}S_{4} = T^{-1}_{2}S_{4}^{-1}T_{3}S_{4}	=T^{-1}_{3}S_{4}^{-1}T^{-1}_{2}S_{4} = I, \\
& (C_{3}C_{2})^{2} = T^{-1}_{3}T^{-1}_{1}(S_{4}C_{3})^{2} = T_{2}(S_{4}C_{2})^{2} = I,                      \\
& T^{-1}_{1}T^{-1}_{2}T_{3}C_{2}(C^{-1}_{3}S_{4})^{2}S_{4} = I,
\end{align}
}
where we use $I$ to denote the identity element of $P4_132$. Note that the three lattice translations $T_1$, $T_2$ and $T_3$ are independent of each other.

The IGG of $P4_{1}32$ is $Z_{2}$. Corresponding to the algebraic relations among lattice symmetric operations, all the constraints of the algebraic PSGs are
{\small
\begin{align}
& G_{C_{2}}(C_{2}(j))G_{C_{2}}(j) = \eta_{2}, \label{eq:psgcon1}                                                                     \\
& G_{C_{3}}(C^{2}_{3}(j))G_{C_{3}}(C_{3}(j))G_{C_{3}}(j) = \eta_{3}, \label{eq:psgcon2}                                              \\
\begin{split}
& G_{T_{1}}^{-1}(S^{3}_{4}(j))G_{S_{4}}(S^{3}_{4}(j))G_{S_{4}}(S^{2}_{4}(j))                                                       \\&\qquad\qquad\qquad\times{}G_{S_{4}}(S_{4}(j))G_{S_{4}}(j) = \eta_{4},
\end{split}\label{eq:psgcon3}\\
& G_{T_{1}}^{-1}(C_{2}T_{1}^{-1}(j))G_{C_{2}}^{-1}(T_{1}^{-1}(j))G_{T_{1}}^{-1}(j)G_{C_{2}}(j) = \eta_{2x}, \label{eq:psgcon4}       \\
& G_{T_{2}}^{-1}(C_{2}T_{3}^{-1}(j))G_{C_{2}}^{-1}(T_{3}^{-1}(j))G_{T_{3}}^{-1}(j)G_{C_{2}}(j) = \eta_{2y}, \label{eq:psgcon5}       \\
& G_{T_{3}}^{-1}(C_{2}T_{2}^{-1}(j))G_{C_{2}}^{-1}(T_{2}^{-1}(j))G_{T_{2}}^{-1}(j)G_{C_{2}}(j) = \eta_{2z}, \label{eq:psgcon6}       \\
& G_{T_{1}}^{-1}(C^{-1}_{3}T_{2}(j))G_{C_{3}}^{-1}(T_{2}(j))G_{T_{2}}(T_{2}(j))G_{C_{3}}(j) = \eta_{3x}, \label{eq:psgcon7}          \\
& G_{T_{2}}^{-1}(C^{-1}_{3}T_{3}(j))G_{C_{3}}^{-1}(T_{3}(j))G_{T_{3}}(T_{3}(j))G_{C_{3}}(j) = \eta_{3y}, \label{eq:psgcon8}          \\
& G_{T_{3}}^{-1}(C^{-1}_{3}T_{1}(j))G_{C_{3}}^{-1}(T_{1}(j))G_{T_{1}}(T_{1}(j))G_{C_{3}}(j) = \eta_{3z}, \label{eq:psgcon9}          \\
& G_{T_{1}}^{-1}(S^{-1}_{4}T_{1}(j))G_{S_{4}}^{-1}(T_{1}(j))G_{T_{1}}(T_{1}(j))G_{S_{4}}(j) = \eta_{4x}, \label{eq:psgcon10}         \\
& G_{T_{2}}^{-1}(S^{-1}_{4}T_{3}(j))G_{S_{4}}^{-1}(T_{3}(j))G_{T_{3}}(T_{3}(j))G_{S_{4}}(j) = \eta_{4y}, \label{eq:psgcon11}         \\
& G_{T_{3}}^{-1}(S^{-1}_{4}T^{-1}_{2}(j))G_{S_{4}}^{-1}(T_{2}^{-1}(j))G_{T_{2}}^{-1}(j)G_{S_{4}}(j) = \eta_{4z}, \label{eq:psgcon12} \\
& G_{C_{3}}(C_{2}(j))G_{C_{2}}(C_{2}C_{3}(j))G_{C_{3}}(C_{3}(j))G_{C_{2}}(j) = \eta_{23}, \label{eq:psgcon13}                        \\
\begin{split}
& G_{T_{2}}(C_{2}(j))G_{S_{4}}(T_{2}^{-1}C_{2}(j))G_{C_{2}}(C_{2}S_{4}(j))                                                         \\&\qquad\times{}G_{S_{4}}(S_{4}(j))G_{C_{2}}(j) = \eta_{24},
\end{split}\label{eq:psgcon14}\\
\begin{split}
& G_{T_{3}}^{-1}(T_{3}C_{3}^{-1}(j))G_{T_{1}}^{-1}(T_{1}T_{3}C_{3}^{-1}(j))G_{S_{4}}(T_{1}T_{3}C_{3}^{-1}(j))                      \\&\qquad\times{}G_{C_{3}}(C_{3}S_{4}(j))G_{S_{4}}(S_{4}(j))G_{C_{3}}(j) = \eta_{34},
\end{split}\label{eq:psgcon15}\\
\begin{split}
& G_{T_{1}}^{-1}(T_{1}S_{4}^{-1}(j))G_{T_{2}}^{-1}(T_{2}T_{1}S_{4}^{-1}(j))G_{T_{3}}(T_{2}T_{1}S_{4}^{-1}(j))                      \\&\times{}G_{C_{2}}(T_{3}^{-1}T_{2}T_{1}S_{4}^{-1}(j))G_{C_{3}}^{-1}(S_{4}C_{3}^{-1}S_{4}(j))\\&\times{}G_{S_{4}}(S_{4}C_{3}^{-1}S_{4}(j))G_{C_{3}}^{-1}(S_{4}(j))G_{S_{4}}(S_{4}(j))G_{S_{4}}(j) = \eta_{234},\label{eq:psgcon16}\\
\end{split}
\end{align}
}
where all $\eta$'s are $Z_{2}$ numbers which take the values of $\pm{}1$, and $j$ may be regarded as combined indices of unit site and sublattice indices, e.g., $j=i\mu$. 
All the possible gauge inequivalent algebraic PSGs can be obtained by solving these coupled constraint equations.

\subsection{Algebraic PSG solutions: Unit cell part}

In this subsection, we prove that all the $P4_{1}32$ PSG elements $G_{U}(i\mu)$ are independent of the unit-cell index $i$. 
Notice there remain lots of gauge redundancy of multiplying each $G_{U}$ with the elements of IGG.
And we can change the sign of $\eta$ by multiplying a PSG generator by $-1$, as long as such a generator appears for odd number times in Eqs.~\eqref{eq:psgcon1}--\eqref{eq:psgcon16}. 
This means that we are able to use freedom to fix some $\eta$'s as follows:
\begin{equation}\label{eq:fixoddG}
\begin{array}{lcr}
G_{C_{2}}{\longrightarrow}-G_{C_{2}} &\Longrightarrow& \eta_{234}=1,  \\
G_{C_{3}}{\longrightarrow}-G_{C_{3}} &\Longrightarrow& \eta_{3}=1,  \\
G_{T_{1}}{\longrightarrow}-G_{T_{1}} &\Longrightarrow& \eta_{4}=1,  \\
G_{T_{2}}{\longrightarrow}-G_{T_{2}} &\Longrightarrow& \eta_{3x}=1,  \\
G_{T_{3}}{\longrightarrow}-G_{T_{3}} &\Longrightarrow& \eta_{4y}=1.  \\
\end{array}
\end{equation}
Notice that the $\eta$'s in Eq.~\eqref{eq:fixoddG} can also be fixed as $-1$, which will lead to the equivalent algebraic PSG solutions.

Before solving the constraints of algebraic PSGs, first we can use the local $U(1)$ gauge redundancy
$G_{U}(i)\rightarrow{}W_{i}G_{U}(i)W^{-1}_{U^{-1}(i)}$ with
$U = T_{1}, T_{2}, T_{3}$ to make
\begin{equation}
\begin{split}
&G_{T_{3}}(0,0,z;\mu) = 1,\\
&G_{T_{2}}(0,y,z;\mu) = 1,\\
&G_{T_{1}}(x,y,z;\mu) = 1,\\
\end{split}
\end{equation}
where we use $(x,y,z;\mu)$ to denote a site in this appendix (and in the main text $i\mu$ is used to be a site index for simplicity).
Then, with $[T_{i}, T_{j}]=0$ for $i,j=1,2,3$, we can obtain
\begin{equation}
\begin{split}
&G_{T_{1}}(x,y,z;\mu) = 1,\\
&G_{T_{2}}(x,y,z;\mu) = (\eta_{xy})^{x},\\
&G_{T_{3}}(x,y,z;\mu) = (\eta_{xz})^{x}(\eta_{yz})^{y}.\\
\end{split}
\end{equation}

Then, the constraints of Eqs.~\eqref{eq:psgcon4}, \eqref{eq:psgcon5}, and \eqref{eq:psgcon6} respectively lead to 
\begin{align}
& G_{C_{2}}(T_{1}(x,y,z;\mu)) = G_{C_{2}}(x,y,z;\mu)\eta_{2x}, \\
\begin{split}
& G_{C_{2}}(T_{3}(x,y,z;\mu)) = G_{C_{2}}(x,y,z;\mu)\eta_{2y} \\&\qquad\qquad\times{}G_{T_{2}}(C_{2}(x,y,z;\mu))(\eta_{xz})^{x} (\eta_{yz})^{y},
\end{split}\\
\begin{split}
& G_{C_{2}}(T_{2}(x,y,z;\mu)) = G_{C_{2}}(x,y,z;\mu)\eta_{2z} \\&\qquad\qquad\times{}G_{T_{3}}(C_{2}(x,y,z;\mu))(\eta_{xy})^{x}.
\end{split}
\end{align}
Above formulas suggest that
\begin{equation}
\begin{split}
&G_{C_{2}}(x,y,z;\mu) =  (\eta_{2x})^{x}(\eta_{2y})^{z}(\eta_{2z})^{y}(\eta_{xz}\eta_{xy})^{x(y+z)}\\&\qquad\times{}{(\eta_{yz})^{yz}}F_{C_{2}}(y,z;\mu)g_{C_{2}}(\mu),
\end{split}
\end{equation}
where
\begin{equation}
F_{C_{2}}(y,z;\mu) = \left\lbrace \begin{array}{ll}
(\eta_{yz})^{y} & \mbox{for\ } \mu =5,\\
(\eta_{xy})^{z}(\eta_{xz})^{y} & \mbox{for\ } \mu =9,\\
1 & \mbox{otherwise}.
\end{array} \right.
\end{equation}
Here we use $g_{U}(\mu)$ to denote $G_{U}(0,0,0;\mu)$ for simplicity. Then the constraint of Eq.~\eqref{eq:psgcon1} gives rise to (consider the site of $3\stackrel{C_{2}}{\longrightarrow}5$)
\begin{equation}\label{eq:eta_xy}
\eta_{2y}\eta_{2z} = 1, \quad \eta_{xz}\eta_{xy} = 1.
\end{equation}
By considering the constraints of Eqs.~\eqref{eq:psgcon10}, \eqref{eq:psgcon11}, and \eqref{eq:psgcon12}, and it is found that
\begin{equation}
G_{S_{4}}(x,y,z;\mu) = (\eta_{4x})^{x}(\eta_{4z})^{y}(\eta_{yz})^{yz}F_{S_{4}}(y,z;\mu)g_{S_{4}}(\mu),
\end{equation}
where
\begin{equation}
F_{S_{4}}(y,z;\mu) = \left\lbrace \begin{array}{ll}
1 & \mbox{for\ } \mu = 1,5,6,\\
(\eta_{yz})^{y}(\eta_{xy})^{y+z} & \mbox{for\ } \mu =7,10,11,\\
(\eta_{yz})^{y} & \mbox{otherwise}.\\
\end{array} \right.
\end{equation}
By considering Eq.~\eqref{eq:psgcon3} associated with the loop of $6\stackrel{S_{4}}{\longrightarrow}9\stackrel{S_{4}}{\longrightarrow}10\stackrel{S_{4}}{\longrightarrow}1\stackrel{S_{4}}{\longrightarrow}6$, we obtain
\begin{equation}
\eta_{xy} = 1, 
\end{equation}
and consequently [see Eq.~\eqref{eq:eta_xy}],
\begin{equation}
\eta_{xz} = 1. 
\end{equation}
Note that here $\eta_4=1$ has been used.
The constraints of Eqs.~\eqref{eq:psgcon7}, \eqref{eq:psgcon8}, and \eqref{eq:psgcon9} lead to 
\begin{equation}
G_{C_{3}}(x,y,z;\mu) = (\eta_{3z})^{x}(\eta_{3y})^{z}(\eta_{yz})^{z(x+y)}g_{C_{3}}(\mu),
\end{equation}
where $\eta_{3x} = 1$ has been used.
By considering Eq.~\eqref{eq:psgcon2} associated with the loop of $1\stackrel{S_{4}}{\longrightarrow}4\stackrel{S_{4}}{\longrightarrow}7\stackrel{S_{4}}{\longrightarrow}1$, we can obtain the relationship between $\eta_{3z}$ and $\eta_{3y}$ as  
\begin{equation}
\eta_{3z}\eta_{3y} = 1.
\end{equation}
Again, by considering Eq.~\eqref{eq:psgcon13} associated with the loop of
$5\stackrel{C_{3}}{\longrightarrow}9\stackrel{C_{2}}{\longrightarrow}9\stackrel{C_{3}}{\longrightarrow}3\stackrel{C_{2}}{\longrightarrow}5$, we can obtain
\begin{equation}
\eta_{yz} = 1,\qquad\eta_{2x}\eta_{2y}\eta_{3z} = 1.
\end{equation}
Based on above analysis, in sum we have
\begin{equation}
G_{T_{1}}=G_{T_{2}}=G_{T_{3}}=1.
\end{equation}
Then, the constraint of Eq.~\eqref{eq:psgcon15} associated with the loop of $1\stackrel{S_{4}}{\longrightarrow}6\stackrel{C_{3}}{\longrightarrow}8\stackrel{S_{4}}{\longrightarrow}7\stackrel{C_{3}}{\longrightarrow}1$ yields
\begin{equation}
\eta_{3z}\eta_{4x}=1.
\end{equation}

These constraints of $\eta_{2x}\eta_{2y}\eta_{3z}=\eta_{2y}\eta_{2z} =\eta_{3z}\eta_{3y} =\eta_{3z}\eta_{4x}=1$ give rise to two independent parameters, namely $\alpha_{1}$ and $\alpha_{2}$. We can also introduce $\alpha_{3}\equiv{}\eta_{4z}$ and then the relations between $\alpha$'s and corresponding $\eta$'s are
\begin{align}
& \alpha_{1} = \eta_{2y} = \eta_{2z}=\pm{}1,                           \\
& \alpha_{2} = \eta_{3y} = \eta_{3z} = \eta_{4x}=\pm{}1,               \\
& \alpha_{3} = \eta_{4z} =\pm{}1, \quad \eta_{2x} = \alpha_{1}\alpha_{2}.
\end{align}
Finally, we can rewrite $G_{U}$'s in terms of independent parameters  $\alpha_{1}$, $\alpha_{2}$, and $\alpha_{3}$ as follows
\begin{align}
& G_{T_{1}}=G_{T_{2}}=G_{T_{3}} =1, \nonumber\\
& G_{C_{2}}(x,y,z;\mu) = \alpha_{1}^{x+y+z}\alpha_{2}^{x}g_{C_{2}}(\mu), \\
& G_{C_{3}}(x,y,z;\mu) = \alpha_{2}^{x+z}g_{C_{3}}(\mu),                 \\
& G_{S_{4}}(x,y,z;\mu) = \alpha_{2}^{x}\alpha_{3}^{y}g_{S_{4}}(\mu).
\end{align}

\subsection{Algebraic PSG solutions: Sublattice part}

In this subsection, the algebraic PSGs for each sublattice are determined by solving the remaining constraint equations. By considering Eq.~\eqref{eq:psgcon3} associated with the sublattices of $6\stackrel{S_{4}}{\longrightarrow}9\stackrel{S_{4}}{\longrightarrow}10\stackrel{S_{4}}{\longrightarrow}1$ as well as the sublattices of  $9\stackrel{S_{4}}{\longrightarrow}10\stackrel{S_{4}}{\longrightarrow}1\stackrel{S_{4}}{\longrightarrow}6$, it is found that
\begin{equation}
\begin{split}
\eta_{4}=1=&g_{S_{4}}(1)g_{S_{4}}(10)g_{S_{4}}(9)g_{S_{4}}(6).\\
=&\alpha_{2}g_{S_{4}}(6)g_{S_{4}}(1)g_{S_{4}}(10)g_{S_{4}}(9),\label{eq:alph2}
\end{split}
\end{equation} 
which indicates that $\alpha_{2} = 1$. 
Similarly, by considering Eq.~\eqref{eq:psgcon14} associated with the sublattices of $4\stackrel{S_{4}}{\longrightarrow}11\stackrel{C_{2}}{\longrightarrow}8\stackrel{S_{4}}{\longrightarrow}7$ as well as the sublattices of  $8\stackrel{S_{4}}{\longrightarrow}7\stackrel{C_{2}}{\longrightarrow}4\stackrel{S_{4}}{\longrightarrow}11$, we can obtain
\begin{equation}
\begin{split}
\eta_{24} = &\alpha_{3}g_{S_{4}}(7)g_{C_{2}}(8)g_{S_{4}}(11)g_{C_{2}}(4)\\
=&\alpha_{1}g_{S_{4}}(11)g_{C_{2}}(4)g_{S_{4}}(7)g_{C_{2}}(8),
\end{split}
\end{equation}
which implies that $\alpha_{1}=\alpha_{3}$. 

Notice that we can still perform the local $U(1)$ gauge transformation as $G_{U}(\mu)\rightarrow{}W_{\mu}G_{U}(\mu)W^{-1}_{U^{-1}(\mu)}$ between two sublattices in order to utilize all the $Z_2$ gauge freedoms. Notice that sublattices $1$, $4$, and $7$ can be transformed by $C_{3}$ as $1\stackrel{C_{3}}{\longrightarrow}4\stackrel{C_{3}}{\longrightarrow}7\stackrel{C_{3}}{\longrightarrow}{}1$, which indicates that for $G_{C_{3}}$ we can always choose a proper gauge to let
\begin{equation}
g_{C_{3}}(4) = g_{C_{3}}(7)=1.
\end{equation}
The condition of $\eta_{3}=1$ indicates that
\begin{equation}
g_{C_{3}}(1)g_{C_{3}}(4)g_{C_{3}}(7) = 1\quad \Longrightarrow\quad{}g_{C_{3}}(1)=1.
\end{equation}
By the same method, by considering the sublattices of $1\stackrel{S_{4}}{\longrightarrow}6\stackrel{S_{4}}{\longrightarrow}9\stackrel{S_{4}}{\longrightarrow}10$, $4\stackrel{S_{4}}{\longrightarrow}11\stackrel{S_{4}}{\longrightarrow}12\stackrel{S_{4}}{\longrightarrow}5$ as well as the sublattices of  $7\stackrel{S_{4}}{\longrightarrow}3\stackrel{S_{4}}{\longrightarrow}2\stackrel{S_{4}}{\longrightarrow}8$, we can always choose a gauge to make
\begin{equation}
g_{S_{4}}(\mu\neq{}1,4,7) = 1.
\end{equation}
So far we have used up all of the gauge redundancy. Then by considering the constraint of Eq.~\eqref{eq:psgcon3} with  $\eta_{4}=1$, we obtain
\begin{equation}
g_{S_{4}}(1) =  g_{S_{4}}(4) = g_{S_{4}}(7) = 1,
\end{equation}
i.e.,
\begin{equation}
g_{S_{4}} = 1.
\end{equation}
In order to solve the constraint of Eq.~\eqref{eq:psgcon13}, we consider some particular loops of $1\stackrel{C_{3}}{\longrightarrow}4\stackrel{C_{2}}{\longrightarrow}7\stackrel{C_{3}}{\longrightarrow}1\stackrel{C_{2}}{\longrightarrow}1$ and $4\stackrel{C_{3}}{\longrightarrow}7\stackrel{C_{2}}{\longrightarrow}4\stackrel{C_{3}}{\longrightarrow}7\stackrel{C_{2}}{\longrightarrow}4$, and it is easy to verify that
\begin{equation}\label{eq:eta23}
\eta_{23} = g_{C_{2}}(1)g_{C_{2}}(7) = g_{C_{2}}^{2}(4),
\end{equation}
where $g_{C_{3}}(1,4,7)=1$ has been used.
Meanwhile, by verifying Eq~\eqref{eq:psgcon1} for each sublattice, we obtain
\begin{equation}\label{eq:eta2}
\begin{split}
\eta_{2} &= g_{C_{2}}^{2}(1) = g_{C_{2}}(2)g_{C_{2}}(12) \\&= \alpha_{1}g_{C_{2}}(3)g_{C_{2}}(5) = g_{C_{2}}(4)g_{C_{2}}(7)\\& = g_{C_{2}}(6)g_{C_{2}}(10) = g_{C_{2}}(8)g_{C_{2}}(11) = \alpha_{1}g_{C_{2}}^{2}(9).
\end{split}
\end{equation}
Similarly, the constrain of Eq.~\eqref{eq:psgcon14} yields
\begin{equation}\label{eq:eta24}
\begin{split}
\eta_{24} &= \alpha_{1}g_{C_{2}}(1)g_{C_{2}}(10) = g_{C_{2}}(2)g_{C_{2}}(11) \\&= g_{C_{2}}(3)g_{C_{2}}(12) = \alpha_{1}g_{C_{2}}(4)g_{C_{2}}(8)\\& = \alpha_{1}g_{C_{2}}(5)g_{C_{2}}(7) = \alpha_{1}g_{C_{2}}(6)g_{C_{2}}(9),
\end{split}
\end{equation}
where $g_{S_{4}}=1$ has been used. Then, Eqs.~\eqref{eq:eta23}, \eqref{eq:eta2}, and \eqref{eq:eta24} lead to 
\begin{align}
& \eta_{2} = \eta_{23},                                   \\
& g_{C_{2}}(1) = g_{C_{2}}(4) = g_{C_{2}}(7),             \\
& g_{C_{2}}(5) = g_{C_{2}}(8) = g_{C_{2}}(10),            \\
& \alpha_{1}g_{C_{2}}(3) = g_{C_{2}}(6) = g_{C_{2}}(11),  \\
& g_{C_{2}}(2) = g_{C_{2}}(9) = \alpha_{1}g_{C_{2}}(12).
\end{align}
Here we study the above equations case by case. For convenience, we introduce a $Z_{2}$ number $\eta_{0}=\pm{}1$. If $\eta_{2} = \eta_{23} = 1$, it is easy to verify that $g_{C_{2}}(1,4,7)=\eta_{0}$,  $g_{C_{2}}(5,8,10)=\alpha_{1}\eta_{0}\eta_{24}$, $g_{C_{2}}(6,11)=\alpha_{1}\eta_{0}\eta_{24}$, $g_{C_{2}}(3)=\eta_{0}\eta_{24}$, $g_{C_{2}}(9,12)=\eta_{0}$, and $g_{C_{2}}(2)=\alpha_{1}\eta_{0}$.
Then, $g_{C_{2}}(2)g_{C_{2}}(12)=\eta_{2}=1$ suggests that $\alpha_{1}=1$. Similarly, if $\eta_{2} = \eta_{23} = -1$, it is also easy to verify that $g_{C_{2}}(1,4,7)=i\eta_{0}$,  $g_{C_{2}}(5,8,10)=-i\alpha_{1}\eta_{0}\eta_{24}$, $g_{C_{2}}(6,11)=-i\alpha_{1}\eta_{0}\eta_{24}$, $g_{C_{2}}(3)=-i\eta_{0}\eta_{24}$, $g_{C_{2}}(9,12)=i\eta_{0}$, and $g_{C_{2}}(2)=i\alpha_{1}\eta_{0}$.
Then, $g_{C_{2}}(2)g_{C_{2}}(12)=\eta_{2}=-1$ suggests that $\alpha_{1}=1$. Overall, it is found that
\begin{equation}
\alpha_{1}=1
\end{equation}
and 
\begin{equation}
\begin{split}
\eta_{2}=1:&\ g_{C_{2}}(1,4,7,9,12,2)=\eta_{0},\\
&\ g_{C_{2}}(5, 8, 10, 6, 11, 3)=\eta_{0}\eta_{24},\\
\eta_{2}=-1:&\ g_{C_{2}}(1,4,7,9,12,2)=i\eta_{0},\\
&\ g_{C_{2}}(5, 8, 10, 6, 11, 3)=-i\eta_{0}\eta_{24}.
\end{split}
\end{equation}

Now we try to solve all the $g_{C_{3}}(\mu)$. First, by considering the constrain of Eq.~\eqref{eq:psgcon2} for each sublattice, we obtain
\begin{equation}
\begin{split}
1 &= g_{C_{3}}(1)g_{C_{3}}(4)g_{C_{3}}(7) = g_{C_{3}}(2)g_{C_{3}}(6)g_{C_{3}}(8)\\&=g_{C_{3}}(5)g_{C_{3}}(9)g_{C_{3}}(3)=g_{C_{3}}(10)g_{C_{3}}(12)g_{C_{3}}(11),
\end{split}
\end{equation}
where $\eta_{3}=1$ has been used.
Furthermore, Eq.~\eqref{eq:psgcon15} yields that
\begin{equation}
\begin{split}
\eta_{34} &= g_{C_{3}}(1)g_{C_{3}}(8) = g_{C_{3}}^{2}(2)\\&=g_{C_{3}}(3)g_{C_{3}}(6)=g_{C_{3}}(4)g_{C_{3}}(10) \\&= g_{C_{3}}(5)g_{C_{3}}(7) = g_{C_{3}}(9)g_{C_{3}}(12) = g_{C_{3}}^{2}(11).
\end{split}
\end{equation}
Similarly, by introducing another $Z_{2}$ number $\eta^\prime_{0}=\pm{}1$, we find that
\begin{equation}
\begin{split}
\eta_{34}=1:&\ g_{C_{3}}(1,4,7,5,8,10)=1,\\ &\ g_{C_{3}}(2,3,6,9,11,12)=\eta^\prime_{0},\\
\eta_{34}=-1:&\ g_{C_{3}}(1,4,7)=1,\ g_{C_{3}}(5,8,10)=-1,\\
&\ g_{C_{3}}(2,3,6,9,11,12)=i\eta^\prime_{0}.
\end{split}
\end{equation}
Notice that $\eta_{24}$ is not a free parameter. By considering the constraint of Eq.~\eqref{eq:psgcon13} associated with the loop of $2\stackrel{C_{3}}{\longrightarrow}6\stackrel{C_{2}}{\longrightarrow}10\stackrel{C_{3}}{\longrightarrow}12\stackrel{C_{2}}{\longrightarrow}2$, one finds that
\begin{equation}
\eta_{24} = \eta_{2}\eta_{34}.
\end{equation}
Up to this point, we summarize all the possible algebraic PSGs as follows:
\begin{subequations}
\begin{align}
\begin{split}\label{eq:case1}
\mbox{Class 1:}&\quad\eta_{2}=1,\ \eta_{34}=1\ \rightarrow{}\eta_{24}=1\\
&g_{C_{2}} = \eta_{0},\quad g_{S_{4}} = 1,\\
&g_{C_{3}}(1,4,7,5,8,10)=1, \\&g_{C_{3}}(2,3,6,9,11,12)=\eta_{0}^\prime,
\end{split}\\
\begin{split}\label{eq:case2}
\mbox{Class 2:}&\quad\eta_{2}=-1,\ \eta_{34}=-1\ \rightarrow{}\eta_{24}=1\\
&g_{C_{2}}(1,4,7,9,12,2)  = i\eta_{0}, \\
&g_{C_{2}}(5,8,10,6,11,3)  = -i\eta_{0},\\
&g_{C_{3}}(1,4,7) = 1,\quad g_{C_{3}}(5,8,10)=-1,\\ &g_{C_{3}}(2,3,6,9,11,12)=i\eta_{0}^\prime,\\
&g_{S_{4}} = 1,
\end{split}\\
\begin{split}\label{eq:case3}
\mbox{Class 3:}&\quad\eta_{2}=1,\ \eta_{34}=-1\ \rightarrow{}\eta_{24}=1\\
&g_{C_{2}} = \eta_{0},\quad g_{S_{4}} = 1,\\
&g_{C_{3}}(1,4,7) = 1,\quad g_{C_{3}}(5,8,10)=-1,\\ &g_{C_{3}}(2,3,6,9,11,12)=i\eta_{0}^\prime,
\end{split}\\
\begin{split}\label{eq:case4}
\mbox{Class 4:}&\quad\eta_{2}=-1,\ \eta_{34}=1\ \rightarrow{}\eta_{24}=1\\
&g_{C_{2}}(1,4,7,9,12,2)  = i\eta_{0}, \\
&g_{C_{2}}(5,8,10,6,11,3)  = -i\eta_{0},\\
&g_{C_{3}}(1,4,7,5,8,10)=1, \\&g_{C_{3}}(2,3,6,9,11,12)=\eta_{0}^\prime,\\
&g_{S_{4}} = 1.
\end{split}
\end{align}
\end{subequations}
However, the algebraic PSGs of Class 3 and Class 4 are forbidden indeed. The constraint in Eq.~\eqref{eq:psgcon16} with $\eta_{234}=1$ yields that
\begin{equation}
g_{C_{2}}(1) = g_{C_{3}}(9)g_{C_{3}}(4),\label{eq:eta234}
\end{equation}
which is only consistent with the PSGs of Class 1 and Class 2. Meanwhile, Eq.~\eqref{eq:eta234} imposes that
\begin{equation}
\eta_{0}=\eta^\prime_{0}.
\end{equation}
Eventually, there are only two free parameters $\eta_{2}$ and $\eta_{0}$, and four kinds of algebraic PSGs that are given in Eq.~\eqref{eq:psg1} and Eq.~\eqref{eq:psg2} in the main text.

\section{Zero-flux sate: Mean-field order parameters}\label{ap:d123}
\begin{figure*}[tbp]
	\includegraphics[width=16.4cm]{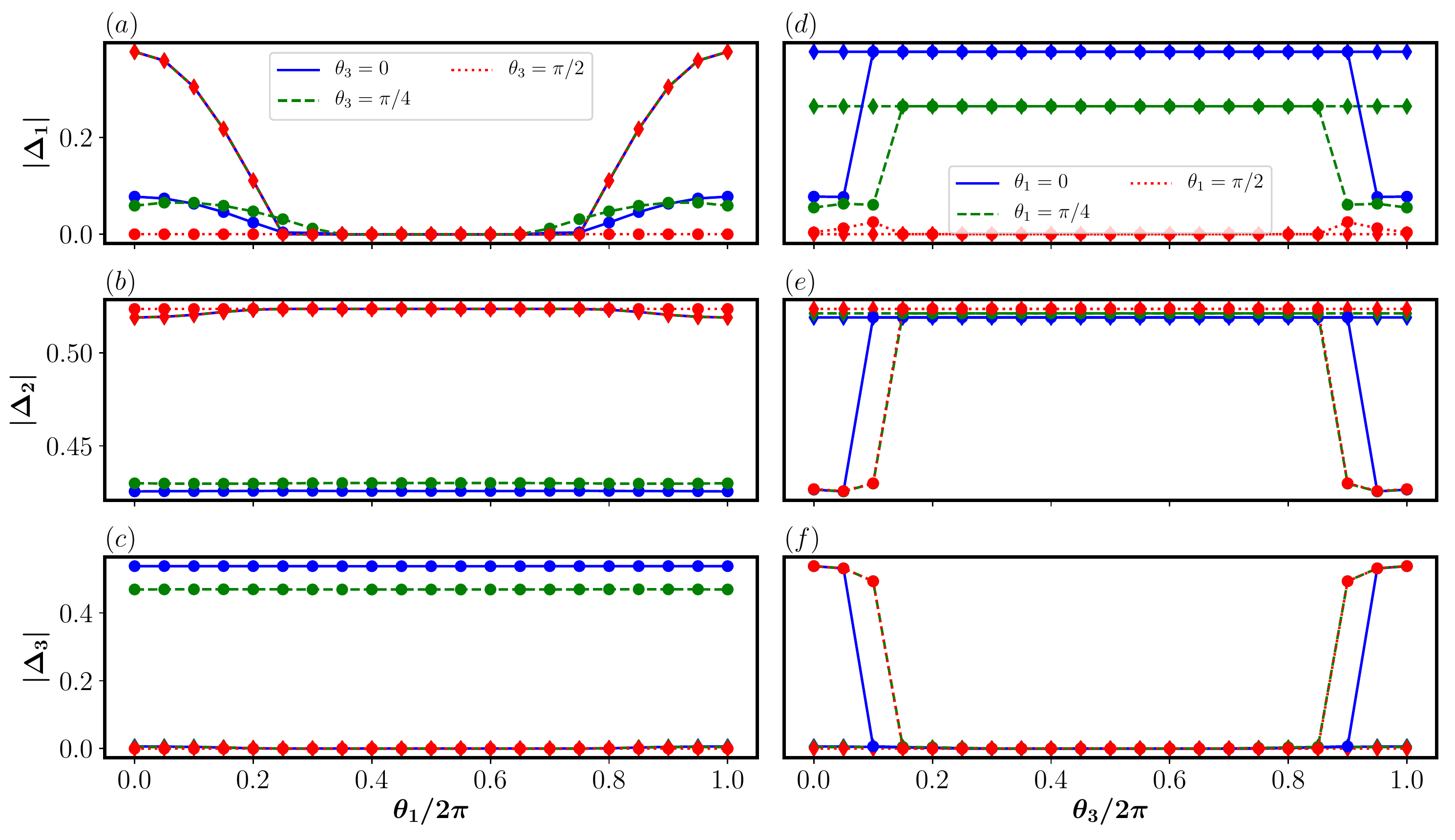}
	\caption{Zero-flux state: order parameters $\Delta_{1,2,3}$ are plotted versus (a) $\theta_1$ and (b) $\theta_3$. Here we set $J_{1}/J_{2}=0.3$, and the diamonds denote $J_{3}/J_{2}=0.3$ (uniform phase) and the circles denote $J_{3}/J_{2}=0.8$ (incommensurate phase).}\label{fig:d123_flux}
\end{figure*}

In this appendix, we provide detailed information for the mean-field order parameters of zero-flux states. As mentioned in the main text, the zero-flux state given in Eq.~\eqref{eq:ansatz2} allows nonzero short-range order parameters on the first three NN bonds, say, $\Delta_{1,2,3}\neq 0$, which is different from the $\pi$-flux state given in Eq.~\eqref{eq:ansatz1} where $\Delta_{1}=\Delta_{3}=0$. As discussed in the main text, we choose $\Delta_{2}$ to be a real number and separate the amplitude and the phase of $\Delta_{1}$ and $\Delta_{3}$ in Eq.~\eqref{eq:theta}. 

For the zero-flux state, the phase ($\theta_{1}$ and $\theta_{3}$) dependence of the ground-state energy $E_{g}$ is illustrated in Fig.~\ref{fig:flux}. Now we would like to demonstrate how the order parameter amplitudes $|\Delta_{1,2,3}|$ change with $\theta_{1}$ and $\theta_{3}$. The results are shown in Fig.~\ref{fig:d123_flux}. It is similar to Fig.~\ref{fig:flux} in that two set of parameters, $(J_{1}/J_{2},J_{3}/J_{2})=(0.3,0.3)$ and $(J_{1}/J_{2},J_{3}/J_{2})=(0.3,0.8)$, are adapted to study the uniform and incommensurate phases respectively.
For $(J_{1}/J_{2},J_{3}/J_{2})=(0.3,0.3)$, $|\Delta_{3}|$ is very small ($<10^{-2}$), so that $E_g$ hardly changes with $\theta_{3}$ as shown in Fig.~\ref{fig:flux}. For $(J_{1}/J_{2},J_{3}/J_{2})=(0.3,0.8)$ and $\theta_3=\pi/2$, $|\Delta_1|$ is almost zero, and therefore the $E_{g}-\theta_{1}$ plot is nearly flat in Fig.~\ref{fig:flux}.

\section{Static spin structure factor}\label{ap:ssf}
In this appendix, we derive the static spin structure factor $S(\bm{q})$ for gapless states and verify that the peaks of $S(\bm{q})$ arise from the spinon condensation. Based on our formalism, the wave vectors $\bm{Q}$ for both the zero-flux state and $\pi$-flux state are discussed. 

Spin operators can be expressed in terms of Schwinger bosons in $k$ space as
\begin{equation}
\begin{split}
&\hat{S}^{x}_{\mu}(\bm{q})=\frac{1}{2N_{u}}\sum_{\bm{k}}\left(b^{\dagger}_{\bm{k}\mu\uparrow}b_{\bm{k+q}\mu\downarrow}+b^{\dagger}_{\bm{k}\mu\downarrow}b_{\bm{k+q}\mu\uparrow}\right),\\
&\hat{S}^{y}_{\mu}(\bm{q})=\frac{-i}{2N_{u}}\sum_{\bm{k}}\left(b^{\dagger}_{\bm{k}\mu\uparrow}b_{\bm{k+q}\mu\downarrow}-b^{\dagger}_{\bm{k}\mu\downarrow}b_{\bm{k+q}\mu\uparrow}\right),\\
&\hat{S}^{z}_{\mu}(\bm{q})=\frac{1}{2N_{u}}\sum_{\bm{k}}\left(b^{\dagger}_{\bm{k}\mu\uparrow}b_{\bm{k+q}\mu\uparrow}-b^{\dagger}_{\bm{k}\mu\downarrow}b_{\bm{k+q}\mu\downarrow}\right).
\end{split}
\end{equation}
The static spin-spin correlation functions are defined as
\begin{equation}
S^{\alpha\beta}_{\mu\nu}(\bm{q})=\langle\hat{S}^{\alpha}_{\mu}(\bm{q})\hat{S}^{\beta}_{\nu}(-\bm{q})\rangle-\langle\hat{S}^{\alpha}_{\mu}(\bm{q})\rangle\langle\hat{S}^{\beta}_{\nu}(-\bm{q})\rangle,
\end{equation}
where $\alpha,\beta=x,y,z$ and
\begin{align}
& S^{\alpha\beta}_{\mu\nu}(\bm{q}) = \delta_{\alpha\beta}S^{\alpha\alpha}_{\mu\nu}(\bm{q}),
\end{align}
due to the spin rotational symmetry for $J_{1}$-$J_{2}$-$J_{3}$ Heisenberg model.

The static spin structure is defined as
\begin{equation}
\mathcal{S}({\bm{q}})=\sum_{\alpha}\sum_{\mu\nu}S^{\alpha\alpha}_{\mu\nu}(\bm{q})=3\sum_{\mu\nu}S^{zz}_{\mu\nu}(\bm{q}),
\end{equation}
where 
\begin{equation}
\begin{split}
S^{zz}_{\mu\nu}(\bm{q})=\frac{1}{4N_{u}^2}&\sum_{\bm{k},\sigma}\Big{\{}
\langle{}b^{\dagger}_{\bm{k}\mu\sigma}b_{\bm{k}\nu\sigma}\rangle\langle{}b_{\bm{k+q}\mu\sigma}b^{\dagger}_{\bm{k+q}\nu\sigma}\rangle\\
-&\langle{}b^{\dagger}_{\bm{k}\mu\sigma}b^{\dagger}_{-\bm{k}\nu\bar{\sigma}}\rangle\langle{}b_{\bm{k+q}\mu\sigma}b_{\bm{-k-q}\nu\bar{\sigma}}\rangle
\Big{\}},
\end{split}
\end{equation}
where $\bar{\sigma}=-\sigma$, e.g., $\bar{\uparrow}=\downarrow$ as well as $\bar{\downarrow}=\uparrow$. 
By Eqs.~\eqref{eq:gamma2b} and \eqref{eq:uv},
at the zero temperature it is easy to verify that
\begin{subequations}
\begin{equation}
\begin{split}
\langle{}b^{\dagger}_{\bm{k}\mu\uparrow}b_{\bm{k}\nu\uparrow}\rangle=&
\sum_{\omega_{\sigma}(\bm{k})\neq{}0}(U^{*}_{\bm{k}})_{\mu\sigma}(U_{\bm{k}})_{\nu\sigma} (v_{\bm{k}\sigma})^{2}\\&\qquad+
\sum_{\omega_{\sigma}(\bm{k})=0}(U^{*}_{\bm{k}})_{\mu\sigma}(U_{\bm{k}})_{\nu\sigma}N_{c},
\end{split}
\end{equation}
\begin{equation}
\begin{split}
\langle{}b^{\dagger}_{-\bm{k}\mu\downarrow}b_{-\bm{k}\nu\downarrow}\rangle=&
\sum_{\omega_{\sigma}(\bm{k})\neq{}0}(V_{\bm{k}})_{\mu\sigma}(V^{*}_{\bm{k}})_{\nu\sigma} (v_{\bm{k}\sigma})^{2}\\&\qquad+
\sum_{\omega_{\sigma}(\bm{k})=0}(V_{\bm{k}})_{\mu\sigma}(V^{*}_{\bm{k}})_{\nu\sigma}N_{c},
\end{split}
\end{equation}
\begin{equation}
\begin{split}
\langle{}b^{\dagger}_{\bm{k}\mu\uparrow}b^{\dagger}_{-\bm{k}\nu\downarrow}\rangle=&
-\sum_{\omega_{\sigma}(\bm{k})\neq{}0}(U^{*}_{\bm{k}})_{\mu\sigma}(V_{\bm{k}})_{\nu\sigma} u_{\bm{k}\sigma}v_{\bm{k}\sigma}\\&\qquad+
\sum_{\omega_{\sigma}(\bm{k})=0}(U^{*}_{\bm{k}})_{\mu\sigma}(V_{\bm{k}})_{\nu\sigma}N_{c},
\end{split}
\end{equation}
\begin{equation}
\begin{split}
\langle{}b^{\dagger}_{-\bm{k}\mu\downarrow}b^{\dagger}_{\bm{k}\nu\uparrow}\rangle=&
-\sum_{\omega_{\sigma}(\bm{k})\neq{}0}(V_{\bm{k}})_{\mu\sigma}(U^{*}_{\bm{k}})_{\nu\sigma} u_{\bm{k}\sigma}v_{\bm{k}\sigma}\\&\qquad+
\sum_{\omega_{\sigma}(\bm{k})=0}(V_{\bm{k}})_{\mu\sigma}(U^{*}_{\bm{k}})_{\nu\sigma}N_{c},
\end{split}
\end{equation}
\end{subequations}
where 
\begin{equation*}
N_{c}=6\alpha{}N_{u}/n_{g}\gg{}1
\end{equation*} 
is the number of condensate spinons per zero mode per flavor and $n_{g}$ is the number of zero modes.

The spinon condensate part significantly contributes to the static spin structure factor. Following the notations in the main text, we use $\bm{K}_{1}$ and $\bm{K}_{2}$ to denote the gapless points, and approximately we obtain
\begin{equation}
\begin{split}\label{eq:conden2sq}
S(\bm{q})\approx&\frac{3N_{c}^{2}}{4N_{u}^{2}}\sum_{...}\Big{\{}
(U_{\bm{K}_{1}}^{*})_{\mu\sigma_{1}}(U_{\bm{K}_{1}})_{\nu\sigma_{1}}(U_{\bm{K}_{2}})_{\mu\sigma_{2}}(U^{*}_{\bm{K}_{2}})_{\nu\sigma_{2}}\\
+&(V_{-\bm{K}_{1}})_{\mu\sigma_{1}}(V^{*}_{-\bm{K}_{1}})_{\nu\sigma_{1}}(V_{-\bm{K}_{2}}^{*})_{\mu\sigma_{2}}(V_{-\bm{K}_{2}})_{\nu\sigma_{2}}\\
-&(U^{*}_{\bm{K}_{1}})_{\mu\sigma_{1}}(V_{\bm{K}_{1}})_{\nu\sigma_{1}}(U_{\bm{K}_{2}})_{\mu\sigma_{2}}(V^{*}_{\bm{K}_{2}})_{\nu\sigma_{2}}\\
-&(V_{-\bm{K}_{1}})_{\mu\sigma_{1}}(U^{*}_{-\bm{K}_{1}})_{\nu\sigma_{1}}(V_{-\bm{K}_{2}}^{*})_{\mu\sigma_{2}}(U_{-\bm{K}_{2}})_{\nu\sigma_{2}}
\Big{\}},\\
\mbox{for\ \ }\bm{q}&=\bm{K}_{1}-\bm{K}_{2},
\end{split}
\end{equation}
and 
\begin{equation}
S({\bm{q}})\ll{}\frac{N_{c}^{2}}{N_{u}^{2}},\quad\mbox{for\ } \bm{q}\neq{}\bm{K}_{1}-\bm{K}_{2},
\end{equation}
where  
\begin{equation*}
\sum_{...}=\sum_{\mu\nu}\sum_{\langle\bm{K}_{1}\bm{K}_{2}\rangle}\sum_{\omega_{\sigma_{1}}(\bm{K}_{1})=0}\sum_{\omega_{\sigma_{2}}(\bm{K}_{2})=0}.
\end{equation*}

It is straightforward to see that the peaks of $S(\bm{q})$ results from the spinon condensation. However, at some certain $\bm{Q}^\prime=\bm{K}_{1}-\bm{K}_{2}$, the corresponding peak may vanish because the terms in Eq.~\eqref{eq:conden2sq} may be canceled by each other, which leads to $S(\bm{Q^\prime})\ll{}N_{c}^{2}/N_{u}^{2}$. In order to see this, first we  express $V_{\bm{k}}$ in terms of $U_{\bm{k}}$ to simplify Eq.~\eqref{eq:conden2sq}. As mentioned in Appendix~\ref{ap:selfcon}, $iA_{\bm{k}}$ is Hermitian, which can be diagonalized with eigenvalues $\tilde{E}_{\bm{k}}$ as 
\begin{equation}
\begin{split}
iA_{\bm{k}}&= \tilde{U}_{\bm{k}}\tilde{E}_{\bm{k}}\tilde{U}^{\dagger}_{\bm{k}}.
\end{split}
\end{equation}
By defining the $12\times{}12$ diagonal matrix $\Upsilon_{\bm{k}}$ ($\mbox{sgn}[\dots]$ is the sign function) as
\begin{equation*}
(\Upsilon_{\bm{k}})_{\mu\nu}=\mbox{sgn}[(\tilde{E}_{\bm{k}})_{\mu}]\delta_{\mu\nu},\  \Upsilon_{\bm{k}}^{2}=I_{12\times{}12},
\end{equation*}
it is easy to verify that
\begin{equation*}
iA_{\bm{k}}=iU_{\bm{k}}E_{\bm{k}}V_{\bm{k}}^\dagger= \tilde{U}_{\bm{k}}\tilde{E}_{\bm{k}}\Upsilon_{\bm{k}}\Upsilon_{\bm{k}}\tilde{U}^{\dagger}_{\bm{k}}=\tilde{U}_{\bm{k}}E_{\bm{k}}\Upsilon_{\bm{k}}\tilde{U}^{\dagger}_{\bm{k}}.
\end{equation*}
where $E_{\bm{k}}$ is the singular values of $A_{\bm{k}}$ defined in Eq.~\eqref{eq:svd}.
Thus, we can always fix a gauge that makes $U_{\bm{k}}=\tilde{U}_{\bm{k}}$, $V_{\bm{k}}=i\tilde{U}_{\bm{k}}\Upsilon_{\bm{k}}$, which suggests that
\begin{equation}
V_{\bm{k}} = i{}U_{\bm{k}}\Upsilon_{\bm{k}}.
\end{equation}
We focus on the case of real mean-field ansatzes (e.g, $A_{\bm{k}}=A^{*}_{-\bm{k}}$), which suggests that 
\begin{equation}
U_{\bm{k}} = U_{-\bm{k}}^{*},\quad{}V_{\bm{k}} = V_{-\bm{k}}^{*}.
\end{equation}
Then Eq.~\eqref{eq:conden2sq} can be simplified as 
\begin{equation}\label{eq:conden2sq_2}
\begin{split}
S({\bm{\bm{Q}}})\approx&\frac{3N_{c}^{2}}{2N_{u}^{2}}\sum_{...}C_{f}(1,2)\\&\times{}(U_{\bm{K}_{1}}^{*})_{\mu\sigma_{1}}(U_{\bm{K}_{1}})_{\nu\sigma_{1}}(U_{\bm{K}_{2}})_{\mu\sigma_{2}}(U^{*}_{\bm{K}_{2}})_{\nu\sigma_{2}},
\end{split}
\end{equation}
where we define
\begin{equation}
C_{f}(1,2)\equiv\left(1-\mbox{sgn}[(\tilde{E}_{\bm{K}_{1}})_{\sigma_{1}}(\tilde{E}_{\bm{K}_{2}})_{\sigma_{2}}]\right),
\end{equation} 
as a factor between gapless points $(\bm{K}_1,\sigma_{1})$ and $(\bm{K}_2,\sigma_{2})$.

A straightforward conclusion can be made that if all the gapless points are not degenerate, then 
\begin{equation} \label{eq:Q=0_1}
S(\bm{Q}=0)\ll{}N_{c}^{2}/N_{u}^{2},
\end{equation} 
due to $C_f(1,1)=0$. On the other hand, because $A_{-\bm{k}}=A_{\bm{k}}^{*}$ and the eigenvalues of $A_{\bm{k}}$ are purely imaginary, it is easy to verify that $C_f(1,\mathcal{T}(1))=2$,  where $\mathcal{T}(1)=(-\bm{K}_{1},\sigma_{1})$ denotes the dual gapless point of $(\bm{K}_{1},\sigma_{1})$ generated by time-reversal operation $\mathcal{T}$. Eventually, we can obtain
\begin{equation}
S(\bm{Q}=2\bm{K})\sim{}N_{c}^{2}/N_{u}^{2}.
\end{equation}

{\em $\pi$-flux state and zero-flux state in uniform phase.} 
The gapless points for the $\pi$-flux state are Eq.~\eqref{eq:pi-K}, and gapless point for the uniform zero-flux state are $\bm{K}=0$ . All these gapless points are time-reversal invariant and so are doubly degenerate. Therefore, for these two states, the peak at $\bm{Q}=2\bm{K}=\bm{0}$ does not vanish, which is protected by time-reversal symmetry.

{\em Zero-flux state in incommensurate phase.}
In this case, the gapless points are at $\bm{K}=K_{0}(\pm{}1,\pm{}1,\pm{}1)$, while the peaks are at Eq.~\eqref{eq:zero-Q} as
\begin{eqnarray}
\bm{Q}&=&2K_0(\pm 1,\pm 1, \pm 1),\, 2K_0(\pm 1,0, 0), \nonumber\\
&& 2K_0(0,\pm 1, 0),\, 2K_0(0,0,\pm 1).  \nonumber
\end{eqnarray}
Notice that all of the gapless points $\bm{K}$ are not degenerate. Therefore, by Eq.~\eqref{eq:Q=0_1}, it is obvious that the peak at $\bm{Q}=0$ vanishes. Meanwhile, it is found numerically that
\begin{equation}
C_f(1,2)=\left\lbrace\begin{array}{rl}
-1,& \bm{K}_{1}\cdot\bm{K}_{2}=K_{0}^{2},\\
1,& \bm{K}_{1}\cdot\bm{K}_{2}=-K_{0}^{2},
\end{array}\right.
\end{equation}
which suggests that the peaks at $2K_0(0,\pm 1, \pm 1)$, $2K_0(\pm 1,0, \pm 1)$, and $ 2K_0(\pm 1,\pm 1, 0)$ also vanish.

\section{Incommensurable boson condensation wave vector $\bm{K}$ for $\kappa_>\kappa_c$}\label{ap:incomm}

The incommensurate phase of the zero-flux state occurs when $\kappa>\kappa_c$, and locates at the bottom right corner of the phase diagram as plotted in Fig.~\ref{fig:phase}, where $J_3>J_{1}$, $J_{3}/J_{2}>\sim 0.62$, and $J_1$ and $J_{3}$ are comparable with each other. In this region, it is numerically found that the magnetically ordered state with boson condensation at $\bm{K}=K_{0}(\pm{}1,\pm{}1,\pm{1})$ has lower energy than the uniform state with boson condensation at $\bm{K}=(0,0,0)$, where $K_{0}/2\pi\approx 0.42-0.5$.

\begin{table}[tbp]
	\caption{The condensation wave vector $\bm{K}=\bm{K}_{[111]}$ in (or nearly in) the $[111]$ direction. Other condensation wave vectors $\bm{K}$ can be obtained by the $P4_{1}32$ symmetry. We set $J_{1}/J_{2}=0.2$ and $J_{3}/J_{2}=0.7\, \mbox{or}\,0.9$.}\label{tab:scaling}
	\renewcommand\arraystretch{1.25}
	\setlength\tabcolsep{0.3cm}
	\begin{tabular}{|c|c|c|}
		\hline
		& $\ J_{3}/J_{2}=0.7$ & $J_{3}/J_{2}=0.9$ \\
		\hline
		$L$ & $\bm{K}_{[111]}/2\pi$ & $\bm{K}_{[111]}/2\pi$                       \\
		\hline
		7  & $(3, 3, 3)/7$              & $(3, 3, 3)/7$                \\
		9  & $(4, 4, 4)/9$           & $(4, 4, 4)/9$              \\
		12  & $(5, 5, 5)/12$          & $(6, 6, 5)/12$                  \\
		12  & $(5, 5, 5)/12$          & $(11, 11, 11)/24$                  \\		27  & $(4, 4, 4)/9$           & $(4, 4, 4)/9$                \\
		28  & $(3, 3, 3)/7$           & $(13, 13, 13)/28$             \\
		30  & $(13, 13, 13)/30$          & $(14, 14, 14)/30$               \\
		48  & $(7, 7, 7)/16$           & $(22, 22, 21)/48$               \\
		63  & $(28, 27, 27)/63$         & $(29, 28, 28)/63$                \\
		69  & $(10, 10, 10)/23$         & $(31, 31, 31)/69$             \\
		\hline
	\end{tabular}
\end{table}

In the incommensurate phase, for finite-size lattices, the condensation wave vectors $\bm{K}$ depend on the ratios $(J_{1}/J_{2}, J_{3}/J_{2})$ and the lattice size $L$, and will converge to finite vectors in the directions of $(\pm 1,\pm 1,\pm 1)$ in the thermodynamic limit $L\to \infty$. In order to see the incommensurability, we set the parameters, $J_{1}/J_{2}=0.2$ and $J_{3}/J_{2}=0.7\, \mbox{or}\,0.9$, and calculate the condensate wave vectors $\bm{K}$ on $L\times L \times L \times 12$ lattices up to $L=69$. The condensate wave vectors in (or nearly in) the [111] direction are list in accordance with $L$ in Table~\ref{tab:scaling}, which is denoted by $\bm{K}_{[111]}$. Other condensation wave vectors can be obtained by $P4_132$ symmetry operations, namely, the $C_4$, $C_3$, and $C_2$ rotations in $k$ space. It is seen that some $\bm{K}_{[111]}$'s in Table~\ref{tab:scaling} are slightly deviated from the direction $(1,1,1)$ because of the finite-size effect. It is expected that the condensation wave vectors $\bm{K}$ will be in the directions of $(\pm 1,\pm 1,\pm 1)$ and $K_0/2\pi$ is an irrational number in the thermodynamic limit $L\to{}\infty$.

\end{appendices}

\bibliography{j1j2j3refer}

\end{document}